\documentclass[%
 reprint,
 superscriptaddress,
 amsmath,
 amssymb,
 amsfonts,
 aps,
]{revtex4-2}

\usepackage[english]{babel}
\addto\captionsenglish{}
\usepackage[utf8x]{inputenc}
\usepackage[T1]{fontenc}
\usepackage{mathrsfs}

\usepackage{float}

\usepackage{graphicx}
\usepackage[colorlinks=true, allcolors=blue]{hyperref}

\usepackage{mathptmx}

\begin{document}

\title{Superconducting Penetration Depth Through a Van Hove Singularity: \texorpdfstring{Sr\textsubscript{2}}\texorpdfstring{RuO\textsubscript{4}} Under Uniaxial Stress}

\author{Eli Mueller}
\affiliation{Stanford Institute for Materials and Energy Sciences, SLAC National Accelerator Laboratory, 2575 Sand Hill Road, Menlo Park, California 94025, USA}
\affiliation{Department of Physics, Stanford University, Stanford, California 94305, USA}

\author{Yusuke Iguchi}

\affiliation{Stanford Institute for Materials and Energy Sciences, SLAC National Accelerator Laboratory, 2575 Sand Hill Road, Menlo Park, California 94025, USA}
\affiliation{Geballe Laboratory for Advanced Materials, Stanford University, Stanford, California 94305, USA}

\author{Fabian Jerzembeck}
\affiliation{Max Planck Institute for the Chemical Physics of Solids, N{\"o}thnitzer Stra{\ss}e 40, Dresden 01187, Germany}

\author{Jorge O. Rodriguez}
\affiliation{Department of Physics, University of Illinois, Urbana, Illinois 61801, USA}

\author{Marisa Romanelli}
\affiliation{Department of Physics, University of Illinois, Urbana, Illinois 61801, USA}

\author{Edgar Abarca-Morales}
\affiliation{Max Planck Institute for the Chemical Physics of Solids, N{\"o}thnitzer Stra{\ss}e 40, Dresden 01187, Germany}

\author{Anastasios Markou}
\affiliation{Max Planck Institute for the Chemical Physics of Solids, N{\"o}thnitzer Stra{\ss}e 40, Dresden 01187, Germany}
\affiliation{Department of Physics, University of Ioannina, 45110 Ioannina, Greece}

\author{Naoki Kikugawa}
\affiliation{National Institute for Materials Science, Tsukuba, Ibaraki 305-0003, Japan}

\author{Dmitry A. Sokolov}
\affiliation{Max Planck Institute for the Chemical Physics of Solids, N{\"o}thnitzer Stra{\ss}e 40, Dresden 01187, Germany}

\author{Gwansuk Oh}
\affiliation{Department of Physics, Graduate School of Science, Kyoto University, Kyoto 606-8502, Japan}
\affiliation{Department of Physics, Pohang University of Science and Technology (POSTECH),  Pohang 790-784, Republic of Korea}

\author{Clifford W. Hicks}
\affiliation{Max Planck Institute for the Chemical Physics of Solids, N{\"o}thnitzer Stra{\ss}e 40, Dresden 01187, Germany}
\affiliation{School of Physics and Astronomy, University of Birmingham, Birmingham B15 2TT, United Kingdom}

\author{Andrew P. Mackenzie}
\affiliation{Max Planck Institute for the Chemical Physics of Solids, N{\"o}thnitzer Stra{\ss}e 40, Dresden 01187, Germany}
\affiliation{Scottish Universities Physics Alliance, School of Physics and Astronomy, University of St. Andrews,St. Andrews KY16 9SS, United Kingdom}

\author{Yoshiteru Maeno}
\affiliation{Department of Physics, Graduate School of Science, Kyoto University, Kyoto 606-8502, Japan}
\affiliation{Toyota Riken - Kyoto University Research Center (TRiKUC), Kyoto University, Kyoto 606-8501, Japan}

\author{Vidya Madhavan}
\affiliation{Department of Physics, University of Illinois, Urbana, Illinois 61801, USA}

\author{Kathryn A. Moler}
\affiliation{Stanford Institute for Materials and Energy Sciences, SLAC National Accelerator Laboratory, 2575 Sand Hill Road, Menlo Park, California 94025, USA}
\affiliation{Department of Physics, Stanford University, Stanford, California 94305, USA}
\affiliation{Geballe Laboratory for Advanced Materials, Stanford University, Stanford, California 94305, USA}

\begin{abstract}
 
In the unconventional superconductor Sr$_2$RuO$_4$, uniaxial stress along the $[100]$ direction tunes the Fermi level through a Van Hove singularity (VHS) in the density of states, causing a strong enhancement of the superconducting critical temperature $T_\textrm{c}$. Here, we report measurements of the London penetration depth $\lambda$ as this tuning is performed. We find that the zero-temperature superfluid density, here defined as $\lambda(0)^{-2}$, increases by $\sim$15\%, with a peak that coincides with the peak in $T_\textrm{c}$. We also find that the low temperature form of $\lambda(T)$ is quadratic over the entire strain range. Using scanning tunneling microscopy, we find that the gap increases from $\Delta_0 \approx 350~\mu$eV in unstressed Sr$_2$RuO$_4$ to $\Delta_0 \approx 600~\mu$eV in a sample strained to near the peak in $T_c$. With a nodal order parameter, an increase in the superconducting gap could bring about an increase in the superfluid density through reduced sensitivity to defects and through reduced non-local effects in the Meissner screening. Our data indicate that tuning to the VHS increases the gap throughout the Brillouin zone, and that non-local effects are likely more important than reduced scattering.

\end{abstract}

\maketitle

 Strontium ruthenate (Sr$_2$RuO$_4$) is a highly studied system in the field of unconventional superconductivity \cite{maeno1994superconductivity,mackenzie2003superconductivity,mackenzie2020personal,mackenzie2017even,maeno2011evaluation,kallin2012,liu2015unconventional,leggett2021symmetry}. The metallic normal state from which superconductivity condenses is one of the simplest and best understood among unconventional superconductors \cite{bergemann2003quasi}, but after nearly three decades of strenuous effort, a complete understanding of the superconducting order parameter remains elusive.

The low-temperature behavior of the London penetration depth $\lambda(T)$ provides information on the existence of nodes in the superconducting gap $\Delta(k)$. The majority of experimental evidence suggests the presence of line nodes~\cite{nishizaki2000changes,deguchi2004gap,kittaka2018searching,izawa2001thermal,suzuki2002universal,hassinger2017vertical}. In the limit of large $\kappa = \lambda/\xi$, where $\lambda$ and $\xi$ are the penetration depth and coherence length respectively, it is well-known that an order parameter with line nodes yields a $T$-linear dependence in the low-temperature form of $\lambda(T)$ for clean samples, and a $T^2$ dependence in the presence of strong impurity scattering~\cite{hirschfeld1993effect}. Measurements of $\lambda(T)$ in Sr$_2$RuO$_4$ have repeatedly shown a $T^2$ form \cite{bonalde2000temperature,ormeno2006electrodynamic,baker2009microwave}. However, it would be odd for the effects of impurity scattering to be strong in Sr$_2$RuO$_4$ because it is one of the cleanest correlated electron materials known, with the best available samples having a mean free path $l$ exceeding 1~$\mu$m \cite{mao2000crystal}. Alternatively, it was suggested in Ref. \cite{bonalde2000temperature} that a $\lambda(T) \sim T^2$ dependence may be a result of non-local electrodynamic effects \cite{kosztin1997nonlocal} because the superconductivity in Sr$_2$RuO$_4$ is only marginally in the local limit with $\kappa \approx$ 2.9 at low temperature \cite{bonalde2000temperature}.

\begin{figure}
 \centerline{\includegraphics[width=\linewidth]{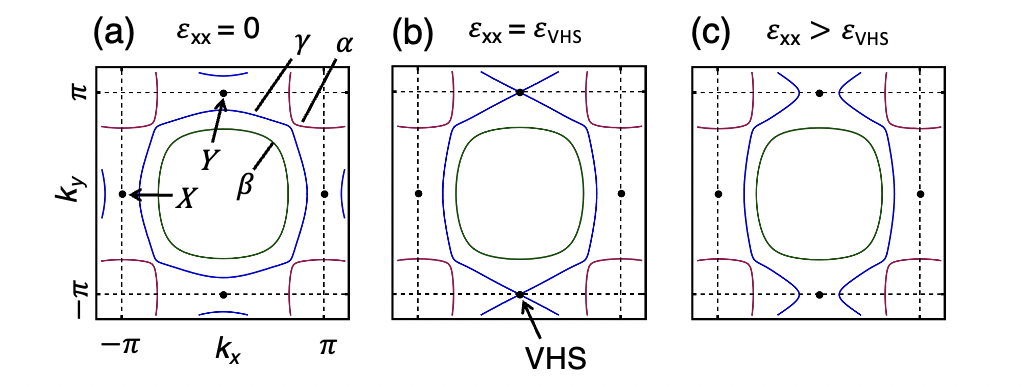}}
  \caption{Quasi-two-dimensional Fermi surface calculations of Sr$_2$RuO$_4$ under representative $\varepsilon_{xx}$ strains \cite{steppke2017strong}. (a) Fermi surface under zero strain, showing the two electron-like bands $\gamma$ and $\beta$ and the hole-like band $\alpha$. Points labeled $X$ and $Y$ indicate high-symmetry points. (b) Fermi surface with the system strain-tuned to the Lifshitz transition. The VHS occurs on the $\gamma$ band at the $Y$ points. (c) Fermi surface for the system strained beyond the Lifshitz transition.}
  \label{fig:fermi_surface}
\end{figure}

The advent of strain-tuning presents a new frontier for studying $\lambda(T)$. Uniaxial stress along the $[100]$ direction drives the $\gamma$ Fermi surface sheet through a Lifshitz transition (see Fig.~\ref{fig:fermi_surface} and Ref.~\cite{sunko2019direct}) and an associated Van Hove singularity (VHS) in the Fermi level density of states. In addition, the superconducting critical temperature $T_{\textrm{c}}$ undergoes a dramatic enhancement from an unstressed value of $T_{\textrm{c}0} = 1.5$~K to a peak of 3.5~K \cite{hicks2014strong,steppke2017strong} at the VHS which is accompanied by a 20-fold increase in the out-of-plane critical field $H_{\textrm{c}2}$ \cite{steppke2017strong,jerzembeck2023upper}. In a weak-coupling picture, the enhanced superconductivity under strain is most intuitively explained by the increased density of states around the VHS \cite{hsu2016manipulating,steppke2017strong}.

There are a few possible hypotheses for how the $T\rightarrow0$ superfluid density $\lambda(0)^{-2}$ might change as the Lifshitz transition is traversed. Theoretical work \cite{mravlje2011coherence} suggests that the quasiparticle mass renormalization on the $\gamma$ sheet is a consequence of the proximity of the Fermi level to the VHS. Tuning to the VHS could cause a further increase in the effective mass $m^*$, thereby decreasing the superfluid density through the relation $\lambda^{-2} = 4\pi n_s e^2/m^*c^2$, where $n_s$ is the number density of superconducting electrons. Alternatively, an increase in the superconducting gap $\Delta_0$ could reduce the effects of impurities and of non-locality in the Meissner screening, leading to an increase in the superfluid density. If the coherence length is strongly reduced by tuning to the Lifshitz transition, then it is also possible that the $T^2$ form of $\lambda(T)-\lambda(0)$ in unstressed Sr$_2$RuO$_4$ would become $T$-linear. We note that a subset of the present authors previously reported scanning superconducting quantum interference device (SQUID) measurements of the penetration depth on uniaxially stressed Sr$_2$RuO$_4$ \cite{mueller2023constraints}.  In those measurements, the focus was to test for an anomaly in the superfluid density associated with a transition below $T_{\textrm{c}}$ \cite{grinenko2021split} and the maximum applied strain $\varepsilon$ was well below the Lifshitz transition. In this letter, we report scanning SQUID measurements of the strain- and temperature-dependent changes of the penetration depth $\lambda(\varepsilon,T)-\lambda(0,0)$ in Sr$_2$RuO$_4$ across the Lifshitz transition. To aid interpretation of these data, we also report measurements by scanning tunneling microscopy (STM) of $\Delta_0$ on Sr$_2$RuO$_4$ strained to near the Lifshitz transition.

\begin{figure}
 \centerline{\includegraphics[width=\linewidth]{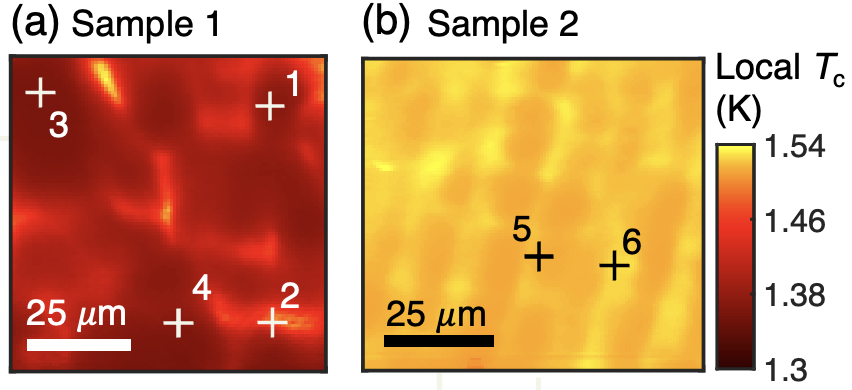}}
  \caption{Local $T_{\textrm{c}}$ maps of (a) sample 1 and (b) sample 2. Markers indicate the positions at which the temperature and strain dependence of the penetration depth was measured.}
  \label{fig:methods_Tc_scan}
\end{figure}

The main components of our SQUID susceptometers \cite{kirtley2016scanning} include a flux-sensitive pickup loop with an inner radius $r_{\textrm{PL}}$ of 1$~\mu$m and a concentric field coil with an inner radius $r_{\textrm{FC}}$ of 2.5$~\mu$m. A low-frequency (893~Hz) alternating current is applied to the field coil, which generates a local field that couples to the SQUID pickup loop. Near a superconducting sample, the Meissner response screens the field from the field coil and reduces the mutual inductance $M$ of the pickup-loop/field-coil pair, which we measure in units of $\Phi_0/\textrm{A}$, where $\Phi_0=h/2e$ is the flux quantum. Recording $M$ as a function of temperature allows us to measure the local $T_{\textrm{c}}$ by measuring the onset of diamagnetism ($M<0$). In the regime where the penetration depth is smaller than $r_{\textrm{FC}}$ and the height of the field coil from the sample surface $z_0$, the measured $M$ can be used to estimate the quantity $\lambda(\varepsilon,T) + z_0$ \cite{kirtley2012scanning}. In our measurement, we fix $z_0$ by bringing the SQUID sensor into light mechanical contact with the sample surface to prevent drift of the SQUID so that changes in $M$ are due to changes in $\lambda(\varepsilon,T)$.

 \begin{figure}
 \centerline{\includegraphics[width=\linewidth]{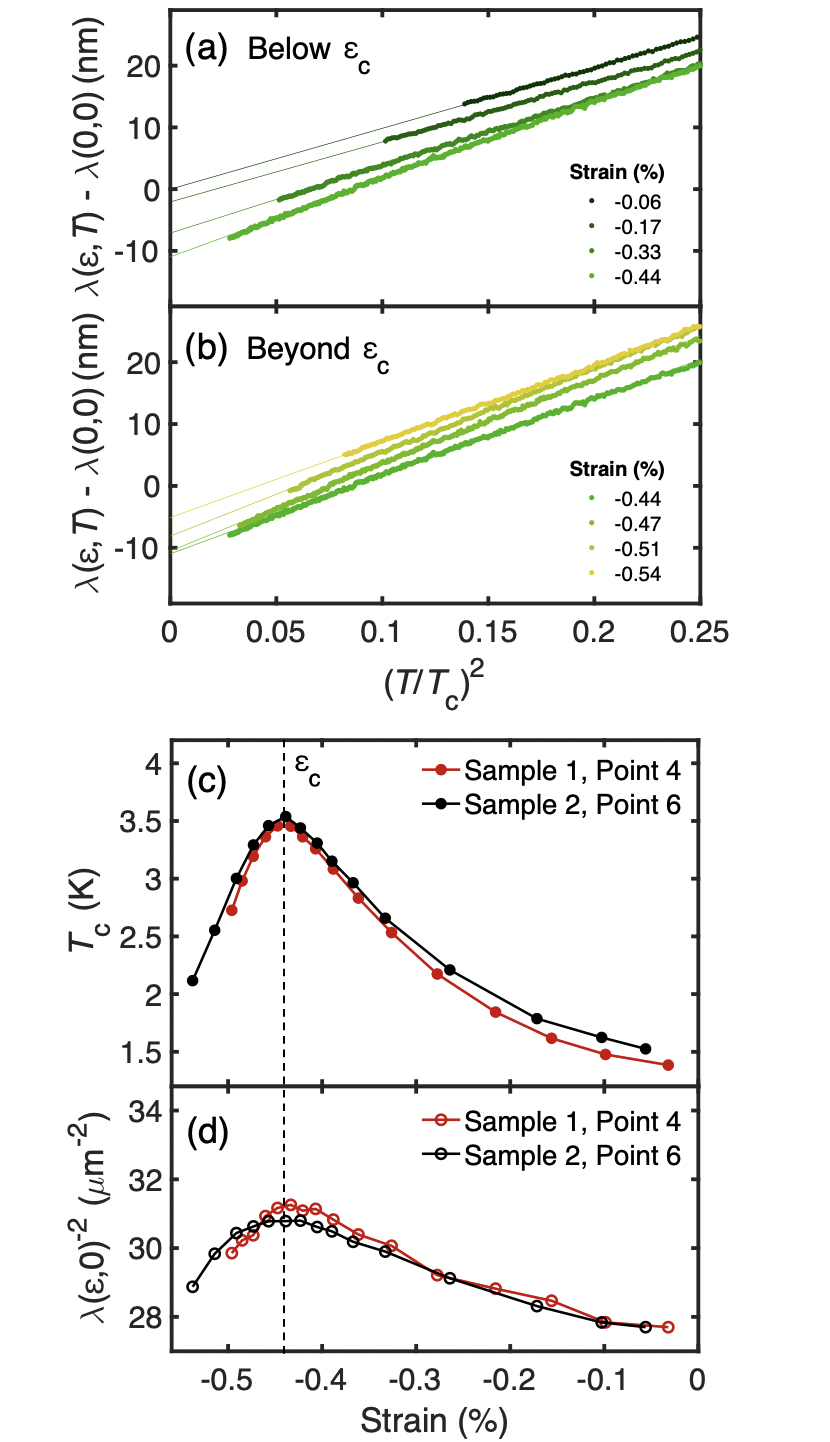}}
  \caption{(a) and (b) Temperature and strain dependence of the change in penetration depth $\lambda(\varepsilon,T)-\lambda(0,0)$ for strains (a) less than $\varepsilon_{\textrm{c}}$ and (b) greater than $\varepsilon_{\textrm{c}}$ for sample 2. Solid lines are $T^2$ fits to the data. We take $\lambda(0,0)$ to be the $T=0$~K intercept of a $\lambda(\varepsilon,T)\sim T^2$ fit to the -0.06\% strain dataset. (c) Strain dependence of local $T_{\textrm{c}}$ measured on sample 1 (red) and 2 (black). (d) Strain dependence of the zero-temperature superfluid density $\lambda(\varepsilon,0)^{-2}$ for sample 1 (red) and sample 2 (black). For each dataset, the zero-temperature superfluid density at the minimum strain is set to $\lambda^{-2} = 27.7$ $\mu$m$^{-2}$ ($\lambda = 190$~nm \cite{bonalde2000temperature}) and we estimate changes in $\lambda(\varepsilon,0)^{-2}$ by extrapolating the $T^2$ fits from (a) and (b) to $T=0$~K.}
  \label{fig:lambda_T}
\end{figure}

For the penetration depth measurements, \textit{in situ} tunable strain was applied using a piezoelectric-driven device similar to that described in previous reports \cite{hicks2014piezoelectric,watson2018micron}. The device contains a capacitive sensor of applied displacement; to convert to strain, we set the strain to be zero at the minimum in $T_\textrm{c}$ and to $\varepsilon_{\textrm{c}} = -0.44$\% at the peak in $T_\textrm{c}$ \cite{barber2019role}. To check reproducibility, we measured two samples cut from separate batches into bars $\approx$2~mm in length. Maps of local $T_{\textrm{c}}$ were obtained near zero strain by taking spatial scans of $M$ at a series of temperatures near the bulk $T_{\textrm{c}}$~\cite{supplemental}. Sample 1  [Fig.~\ref{fig:methods_Tc_scan}(a)] shows a lower overall $T_{\textrm{c}}$ of 1.36--1.4~K with a few small pockets of local $T_{\textrm{c}}$ values as high as 1.5~K. Sample 2 [Fig.~\ref{fig:methods_Tc_scan}(b)] is more homogeneous and shows higher overall $T_{\textrm{c}}$ values of 1.51--1.54~K. The apparent stripe pattern in the local $T_{\textrm{c}}$ map of sample 2 is most likely due to periodic changes in the solidification condition during sample growth \cite{nishizaki1996effect}\footnote{$T_{\textrm{c}}$ has been shown to be relatively stable to changes in oxygen partial pressure annealing conditions \cite{nishizaki1996effect}. Therefore, the stripe features in local $T_{\textrm{c}}$ most likely correspond to ruthenium deficiencies that periodically accumulate during sample growth.}.

 \begin{figure}
 \centerline{\includegraphics[width=\linewidth]{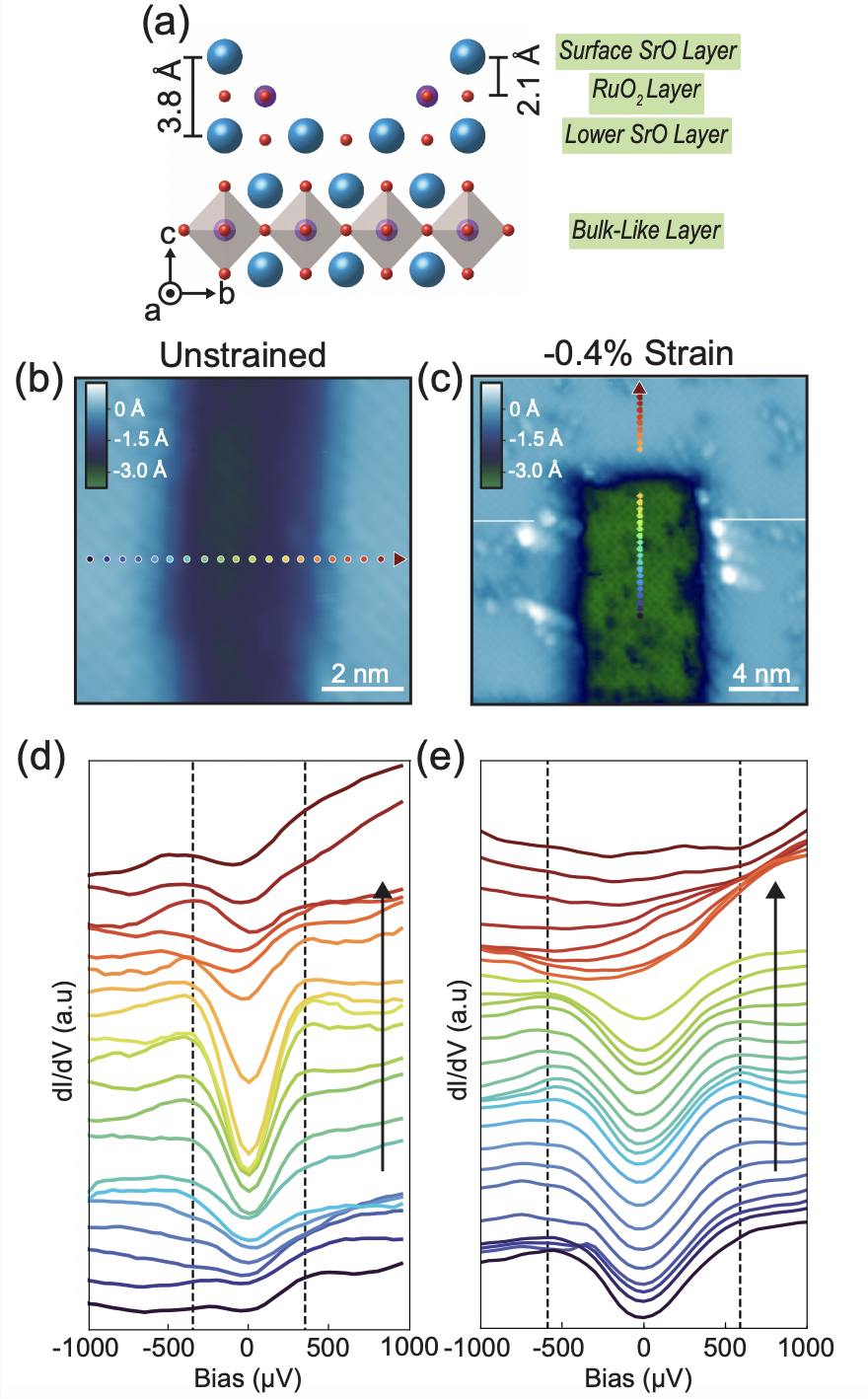}}
  \caption{Superconducting gap under the surface SrO layer in Sr$_2$RuO$_4$. (a) Schematic diagram of the crystal structure of Sr$_2$RuO$_4$ showing subsurface layers (b) Topographic image of a defect (dark blue area) on an unstrained sample exposing the RuO$_2$ layer at a depth of $\approx 2.1$~\r{A} (c) Topographic image of a defect (green area) on a strained sample exposing the lower SrO layer at a depth of $\approx 3.4$~\r{A}. See Ref. \cite{supplemental} for details for determining the depth of the cleavage defects. The topographic depth indicated by the colorbar is measured relative to the surface SrO layer (light blue area) (d) Waterfall plot of $dI/dV$ spectroscopy data measured at the points indicated in (b). (e) Waterfall plot of $dI/dV$ spectroscopy data measured at the points indicated in (c). The superconducting gap is determined from the distance between the pair of dashed lines in panels (d) and (e). a.u.: arbitrary units.}
  \label{fig:STM_strained_unstrained}
\end{figure}




 Figures~\ref{fig:lambda_T}(a) and (b), show the temperature and strain dependence of the change in penetration depth $\lambda(\varepsilon,T)-\lambda(0,0)$ plotted against $(T/T_{\textrm{c}})^2$. For fixed $T/T_{\textrm{c}}$, the curves show a clear reduction in penetration depth with increasing strain below $\varepsilon_{\textrm{c}}$ [Fig.~\ref{fig:lambda_T}(a)] and a subsequent increase in penetration depth for strains beyond $\varepsilon_{\textrm{c}}$ [Fig.~\ref{fig:lambda_T}(b)]. The $T^2$ fits (solid lines) agree well with the data for $T<0.5 T_{\textrm{c}}$ over the entire strain range. 
 

In Fig.~\ref{fig:lambda_T}(c), the $T_{\textrm{c}}$ at two selected points is plotted against applied strain. The local $T_{\textrm{c}}$ for each value of strain was determined from $M(T)$ by the temperature at which the maximum in $dM/dT$ occurs \cite{supplemental}. The $T_{\textrm{c}}$ versus strain curves for sample 1 and sample 2 
 approximately track each other; however, the difference between $T_\textrm{c}$ of the two samples shrinks as $T_\textrm{c}$ increases, consistent with the general expectation that higher $T_\textrm{c}$ corresponds to reduced sensitivity to defect scattering. Figure \ref{fig:lambda_T}(d) shows the strain dependence of the zero-temperature superfluid density $\lambda(\varepsilon,0)^{-2}$.  Under increasing compression, $\lambda(\varepsilon,0)^{-2}$ increases monotonically up to $\varepsilon_{\textrm{c}}$, where it reaches an approximately 15\% enhancement, then decreases for strains beyond $\varepsilon_{\textrm{c}}$.

For the STM measurements, due to the restricted sample space, strain was applied using a differential thermal expansion cell as in Ref.~\cite{sunko2019direct}. We estimate that the applied strain in the sample was 0.4$\pm$0.1\% using digital image correlation of the sample platform at low temperature; additional details are given in Ref.~\cite{supplemental}. Data from the STM measurements are shown in Fig.~\ref{fig:STM_strained_unstrained}. As in previous works \cite{marques2021magnetic,wang2017quasiparticle,kambara2006scanning,barker2003stm,kreisel2021quasi}, we do not observe a superconducting gap on the top strontium oxide (SrO) surface, possibly because of the surface reconstruction\cite{matzdorf2000ferromagnetism,veenstra2013arpes,morales2023hierarchy}. However, we do observe a superconducting gap inside several nanometer-scale trenches, most probably created when the sample was cleaved.  In these trenches both the surface SrO layer and the RuO$_2$ layer directly below it are removed, thereby exposing the lower SrO layer which provides a window into the bulk properties [see Fig.~\ref{fig:STM_strained_unstrained}(a)].


Figures~\ref{fig:STM_strained_unstrained}(b) and (c) show typical topographies as well as spectroscopic linecutes [Fig.~\ref{fig:STM_strained_unstrained}(d) and Fig.~\ref{fig:STM_strained_unstrained}(e)] in these trenches in an unstrained and strained sample, respectively. In the unstrained sample [Fig.~\ref{fig:STM_strained_unstrained}(d)], the spectra inside the trench show a well-defined gap with 2$\Delta_0$ equal to approximately 700~$\mu$V, as denoted by the dashed lines marking the position of the coherence peaks. We note that this value of 2$\Delta_0$ is consistent with previous measurements on unstrained Sr$_2$RuO$_4$~\cite{firmo2013evidence,sharma2020momentum}. In contrast, as illustrated by the dashed lines in Fig.~\ref{fig:STM_strained_unstrained}(e), on the strained sample, we find a much larger value for 2$\Delta_0$ of approximately 1.2~mV. We note that the gap measured by STM is likely to be dominated by the $\alpha$ and $\beta$ sheets. As discussed in Ref.~\cite{firmo2013evidence}, the $\alpha$ and $\beta$ sheets are dominated by $xz$ and $yz$ orbital weight and the tunneling conductance is likely to be much higher for the $xz$ and $yz$ orbitals than the $xy$ orbital due to their greater extent along the $z$ axis.

 \begin{figure}
 \centerline{\includegraphics[width=\linewidth]{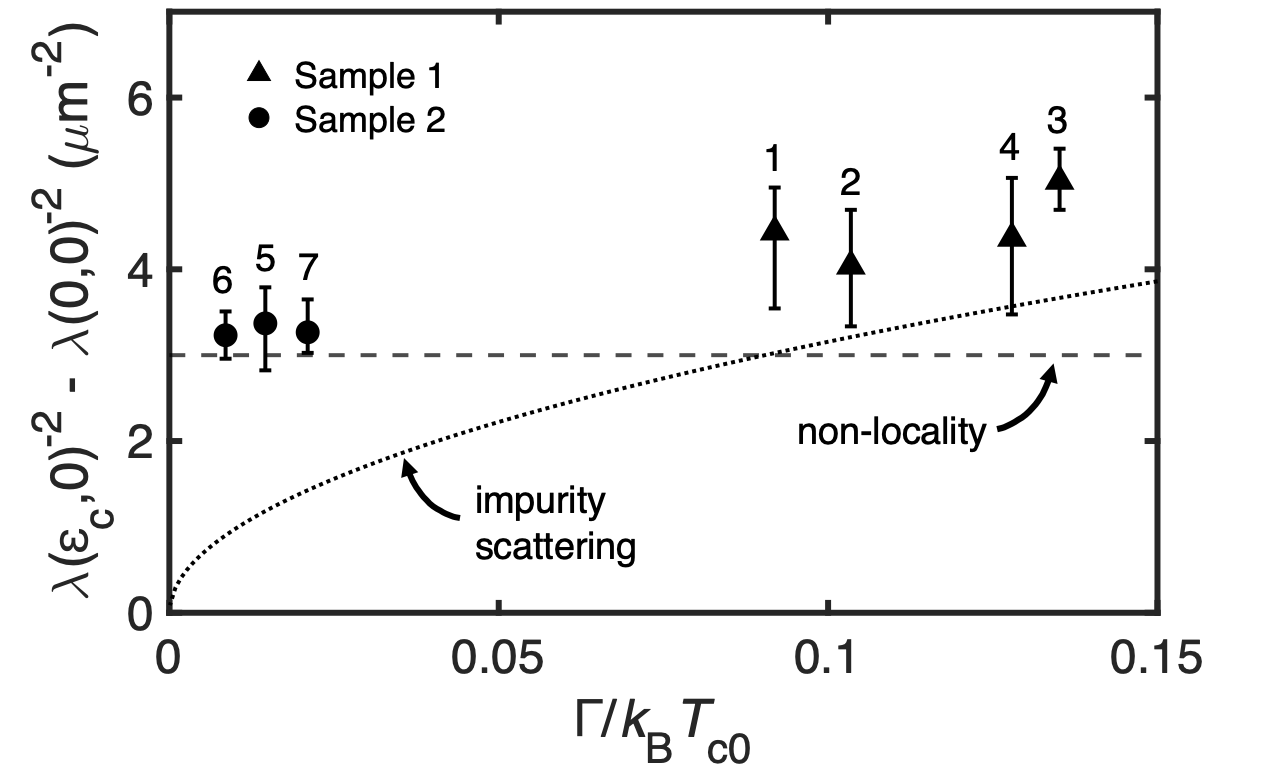}}
  \caption{Superfluid density enhancement at $\varepsilon_{\textrm{c}}$ plotted against the scattering rate $\Gamma$ determined from the unstrained $T_c$ as described in the text and normalized by $k_{\textrm{B}}T_{c0}$. The labels at each marker correspond to the location on the samples indicated in Fig.~\ref{fig:methods_Tc_scan}(a) and (b). Point 7 was measured outside the scan range of Fig.~\ref{fig:methods_Tc_scan}(b). The error bars indicate the range of values observed over several measurement cycles and the data points represent the average value. The dotted line shows the prediction from reduced sensitivity to impurities, and the dashed line from reduced non-locality, both under an assumption that $\Delta_0$ increases in proportion to $T_{\textrm{c}}$.}
  \label{fig:deltans_v_Tcmin}
\end{figure}



Importantly for discussion of the superfluid density, the gap on the $\alpha$ and $\beta$ sheets is well away from the Van Hove point in $k$-space (points labeled $Y$ in Fig.~\ref{fig:fermi_surface}); due to the low Fermi velocity, carriers near the Van Hove point will not contribute strongly to the superfluid density.  We therefore restrict our analysis of the superfluid density results to regions of the Fermi surface away from the Van Hove point. An enhanced $m^*$ should increase $\lambda(0)$, thereby reducing the superfluid density. In contrast, our results show an increase in superfluid density, suggesting that any enhancement in $m^*$ near the Lifshitz transition is too small to have a dominant effect. Angle-resolved photoemission spectroscopy (ARPES) data show that there is no substantial change in band renormalization as the $\gamma$ sheet is tuned to the Lifshitz transition by substitution \cite{shen2007evolution}, and ARPES data under uniaxial stress show, similarly, no dramatic change \cite{sunko2019direct}. Therefore, we suppose that $m^*$ on sections of the Fermi surface away from the Van Hove point remains approximately constant as stress is applied.  

We suppose further that away from the Van Hove point, the gap increases in proportion to $T_{\textrm{c}}$ (the gap is likely substantial at the Van Hove point \cite{li2022elastocaloric}, but the superfluid density is not sensitive to this due to the low Fermi velocity).  For a nodal superconductor, an enhanced gap can lead to an increased superfluid density in two ways: (1) reduced sensitivity to impurity scattering and (2) reduced non-local effects in the Meissner screening. For the case of impurities in a $d$-wave superconductor \cite{hirschfeld1993effect}, it was shown that $\lambda(0)$ increases from its value in the clean, local limit $\lambda_0$ by an amount that depends on the magnitude of the superconducting gap: $\lambda(0)-\lambda_0\sim\lambda_0\sqrt{\Gamma/\Delta_0}$, where $\Gamma$ is the nonmagnetic impurity scattering rate. Studies of $T_{\textrm{c}}$ on samples with varying levels of impurities \cite{mackenzie1998extremely,mao1999suppression,kikugawa2002non,kikugawa2004rigid} show a suppression of $T_{\textrm{c}}$ with increasing impurity concentration that follows an Abrikosov-Gorkov (AG) relation,  $\ln(T_{\textrm{c}0}/T_{\textrm{c}}) =  \psi(1/2 + \Gamma/2\pi k_{\textrm{B}}T_{\textrm{c}}) - \psi(1/2)$
\cite{abrikosov1960contribution,hirschfeld1988consequences}, where $\psi$ is the digamma function and $T_{\textrm{c0}}$ is the clean-limit $T_{\textrm{c}}$. Therefore, taking $T_{\textrm{c}}$ as a measure of $\Gamma$, we would expect a greater enhancement of the superfluid density at the Lifshitz transition for samples that have lower unstressed $T_{\text{c}}$.

In Fig.~\ref{fig:deltans_v_Tcmin}, we show the increase in superfluid density at $\varepsilon_\textrm{c}$ for each point measured on the samples (see Fig.~\ref{fig:methods_Tc_scan}) plotted against $\Gamma/k_{\textrm{B}}T_{\textrm{c0}}$, where $\Gamma$ is determined from the AG relation applied to $T_{\textrm{c}}(\varepsilon=0)$. Measurements on sample 1, which has lower $T_\text{c}(\varepsilon=0)$, indeed show a larger increase in superfluid density at the Lifshitz transition. However, the trend of the data does not match the expectation from impurity effects alone: the increase in superfluid density in sample 2 remains large, even though its $T_\text{c}(\varepsilon=0)$ is very close to the clean-limit value.

At sufficiently low temperature, non-local effects in the Meissner screening are expected to become important in all nodal superconductors because the coherence length, defined as $\xi(k)= \hbar v_{\textrm{F}}(k)/\pi\Delta(k)$, diverges along the nodal directions such that $\xi(k)>\lambda_0$ is satisfied near the nodes \cite{kosztin1997nonlocal}. Similar to the effect of impurities, the low-temperature behaviour of $\lambda(T)$ becomes quadratic and $\lambda(0)$ increases from $\lambda_0$. In a $d$-wave superconductor \cite{kosztin1997nonlocal}, the correction to $\lambda(0)$ can be estimated as $\lambda(0)-\lambda_\textrm{0} = \lambda_\textrm{0}\pi\sqrt{2}/16\kappa$. Taking $\lambda_0 = 190$~nm and a zero-temperature coherence length of $\xi_0 =$66~nm \cite{mackenzie2003superconductivity}, the non-local correction to $\lambda(0)$ is approximately 10\% of $\lambda_{\textrm{0}}$ in unstressed Sr$_2$RuO$_4$. $T_{\text{c}}$ increases by a factor of 2.3 between $\varepsilon=0$ and the Lifshitz transition. If we take $\kappa$ to increase by this same factor, we find that $\lambda(0)$ decreases by $\approx 10$~nm; $\lambda(0)^{-2}$ increases by $\approx 3$~$\mu$m$^{-2}$. This increase is shown by the dashed line in Fig. \ref{fig:deltans_v_Tcmin} and is consistent with the superfluid enhancement observed on sample 2. We note that the twenty-fold increase in $H_{\textrm{c2}}$ \cite{steppke2017strong,jerzembeck2023upper} indicates a greater decrease in $\xi_0$ by a factor of $\sqrt{20}$. However, the highly anisotropic electronic structure of Sr$_2$RuO$_4$ tuned to the Lifshitz transition should be recalled: this increase in $H_{\textrm{c2}}$ could be driven by carriers in the immediate vicinity of the Van Hove point. The value of the penetration depth measurements is that they provide information on how portions of the Fermi surface away from the Van Hove point, where the Fermi velocity is larger, respond to the tuning to the Lifshitz transition.

In summary, the zero-temperature superfluid density in Sr$_2$RuO$_4$ undergoes a $\sim$15\% enhancement as the system is strain-tuned through the VHS, and the penetration depth shows a $\lambda(T)\sim T^2$ dependence throughout the Lifshitz transition. In addition, the superconducting gap increases with strain going from $\Delta_0 \approx 350~\mu$eV in an unstrained sample to $\Delta_0 \approx 600~\mu$eV in a sample strained to near the Lifshitz transition. Our analysis indicates that the increase in superfluid density with tuning to the Lifshitz transition is primarily driven by a decrease in non-local effects in the Meissner screening. We note that Ref.~\cite{landaeta2023nonlocal} reports measurements of $H_{\textrm{c1}}$ in unstressed Sr$_2$RuO$_4$ that also highlight the importance of non-local effects. Both the STM and the penetration depth measurements, which are sensitive primarily to regions of Fermi surface away from the Van Hove point, show that tuning to the Lifshitz transition increases the gap throughout the Brillouin zone, an observation that might be a useful clue on the pairing mechanism in Sr$_2$RuO$_4$.

\newpage

This work was supported by the Department of Energy, Office of Basic Energy Sciences, Division of Materials Sciences and Engineering, under contract DE-AC02-76SF00515. N. K. is supported by JSPS KAKENHI (No. JP18K04715, No. JP21H01033, and No. JP22K19093). YM was supported by JSPS KAKENHI (Nos. JP15H05851 and JP22H01168). GSO was supported by JSPS KAKENHI (Nos. JP15H05851, JP15K21717) during his stay at Kyoto University. We thank Gwansuk Oh for his contribution to the crystal growth. F.J., A.P.M. and C.W.H. acknowledge the support of the Max Planck Society and the German Research Foundation (TRR288-422213477 ELASTO-Q-MAT, Project A10).  Research in Dresden benefits from the environment provided by the DFG Cluster of Excellence (ct.qmat EXC 2147, Project 390858490. C.W.H. acknowledges support from the Engineering and Physical Sciences Research Council (U.K.) (EP/X01245X/1). This work was supported by the Center for Quantum Sensing and Quantum Materials, an Energy Frontier Research
Center funded by the U. S. Department of Energy, Office of Science, Basic Energy Sciences under Award DE-SC0021238. We acknowledge useful conversations with J. Landaeta and E. Hassinger.

\bibliography{SRO}

\begin{thebibliography}{56}%
\makeatletter
\providecommand \@ifxundefined [1]{%
 \@ifx{#1\undefined}
}%
\providecommand \@ifnum [1]{%
 \ifnum #1\expandafter \@firstoftwo
 \else \expandafter \@secondoftwo
 \fi
}%
\providecommand \@ifx [1]{%
 \ifx #1\expandafter \@firstoftwo
 \else \expandafter \@secondoftwo
 \fi
}%
\providecommand \natexlab [1]{#1}%
\providecommand \enquote  [1]{``#1''}%
\providecommand \bibnamefont  [1]{#1}%
\providecommand \bibfnamefont [1]{#1}%
\providecommand \citenamefont [1]{#1}%
\providecommand \href@noop [0]{\@secondoftwo}%
\providecommand \href [0]{\begingroup \@sanitize@url \@href}%
\providecommand \@href[1]{\@@startlink{#1}\@@href}%
\providecommand \@@href[1]{\endgroup#1\@@endlink}%
\providecommand \@sanitize@url [0]{\catcode `\\12\catcode `\$12\catcode
  `\&12\catcode `\#12\catcode `\^12\catcode `\_12\catcode `\%12\relax}%
\providecommand \@@startlink[1]{}%
\providecommand \@@endlink[0]{}%
\providecommand \url  [0]{\begingroup\@sanitize@url \@url }%
\providecommand \@url [1]{\endgroup\@href {#1}{\urlprefix }}%
\providecommand \urlprefix  [0]{URL }%
\providecommand \Eprint [0]{\href }%
\providecommand \doibase [0]{https://doi.org/}%
\providecommand \selectlanguage [0]{\@gobble}%
\providecommand \bibinfo  [0]{\@secondoftwo}%
\providecommand \bibfield  [0]{\@secondoftwo}%
\providecommand \translation [1]{[#1]}%
\providecommand \BibitemOpen [0]{}%
\providecommand \bibitemStop [0]{}%
\providecommand \bibitemNoStop [0]{.\EOS\space}%
\providecommand \EOS [0]{\spacefactor3000\relax}%
\providecommand \BibitemShut  [1]{\csname bibitem#1\endcsname}%
\let\auto@bib@innerbib\@empty
\bibitem [{\citenamefont {Maeno}\ \emph {et~al.}(1994)\citenamefont {Maeno},
  \citenamefont {Hashimoto}, \citenamefont {Yoshida}, \citenamefont
  {Nishizaki}, \citenamefont {Fujita}, \citenamefont {Bednorz},\ and\
  \citenamefont {Lichtenberg}}]{maeno1994superconductivity}%
  \BibitemOpen
  \bibfield  {author} {\bibinfo {author} {\bibfnamefont {Y.}~\bibnamefont
  {Maeno}}, \bibinfo {author} {\bibfnamefont {H.}~\bibnamefont {Hashimoto}},
  \bibinfo {author} {\bibfnamefont {K.}~\bibnamefont {Yoshida}}, \bibinfo
  {author} {\bibfnamefont {S.}~\bibnamefont {Nishizaki}}, \bibinfo {author}
  {\bibfnamefont {T.}~\bibnamefont {Fujita}}, \bibinfo {author} {\bibfnamefont
  {J.}~\bibnamefont {Bednorz}},\ and\ \bibinfo {author} {\bibfnamefont
  {F.}~\bibnamefont {Lichtenberg}},\ }\bibfield  {title} {\bibinfo {title}
  {Superconductivity in a layered perovskite without copper},\ }\href@noop {}
  {\bibfield  {journal} {\bibinfo  {journal} {Nature}\ }\textbf {\bibinfo
  {volume} {372}},\ \bibinfo {pages} {532} (\bibinfo {year}
  {1994})}\BibitemShut {NoStop}%
\bibitem [{\citenamefont {Mackenzie}\ and\ \citenamefont
  {Maeno}(2003)}]{mackenzie2003superconductivity}%
  \BibitemOpen
  \bibfield  {author} {\bibinfo {author} {\bibfnamefont {A.~P.}\ \bibnamefont
  {Mackenzie}}\ and\ \bibinfo {author} {\bibfnamefont {Y.}~\bibnamefont
  {Maeno}},\ }\bibfield  {title} {\bibinfo {title} {The superconductivity of
  {Sr$_2$RuO$_4$} and the physics of spin-triplet pairing},\ }\href@noop {}
  {\bibfield  {journal} {\bibinfo  {journal} {Rev. Mod. Phys.}\ }\textbf
  {\bibinfo {volume} {75}},\ \bibinfo {pages} {657} (\bibinfo {year}
  {2003})}\BibitemShut {NoStop}%
\bibitem [{\citenamefont {Mackenzie}(2020)}]{mackenzie2020personal}%
  \BibitemOpen
  \bibfield  {author} {\bibinfo {author} {\bibfnamefont {A.~P.}\ \bibnamefont
  {Mackenzie}},\ }\bibfield  {title} {\bibinfo {title} {A personal perspective
  on the unconventional superconductivity of {Sr$_2$RuO$_4$}},\ }\href@noop {}
  {\bibfield  {journal} {\bibinfo  {journal} {J. Supercond. Novel Mag.}\
  }\textbf {\bibinfo {volume} {33}},\ \bibinfo {pages} {177} (\bibinfo {year}
  {2020})}\BibitemShut {NoStop}%
\bibitem [{\citenamefont {Mackenzie}\ \emph {et~al.}(2017)\citenamefont
  {Mackenzie}, \citenamefont {Scaffidi}, \citenamefont {Hicks},\ and\
  \citenamefont {Maeno}}]{mackenzie2017even}%
  \BibitemOpen
  \bibfield  {author} {\bibinfo {author} {\bibfnamefont {A.~P.}\ \bibnamefont
  {Mackenzie}}, \bibinfo {author} {\bibfnamefont {T.}~\bibnamefont {Scaffidi}},
  \bibinfo {author} {\bibfnamefont {C.~W.}\ \bibnamefont {Hicks}},\ and\
  \bibinfo {author} {\bibfnamefont {Y.}~\bibnamefont {Maeno}},\ }\bibfield
  {title} {\bibinfo {title} {Even odder after twenty-three years: the
  superconducting order parameter puzzle of {Sr$_2$RuO$_4$}},\ }\href@noop {}
  {\bibfield  {journal} {\bibinfo  {journal} {npj Quantum Mater.}\ }\textbf
  {\bibinfo {volume} {2}},\ \bibinfo {pages} {1} (\bibinfo {year}
  {2017})}\BibitemShut {NoStop}%
\bibitem [{\citenamefont {Maeno}\ \emph {et~al.}(2011)\citenamefont {Maeno},
  \citenamefont {Kittaka}, \citenamefont {Nomura}, \citenamefont {Yonezawa},\
  and\ \citenamefont {Ishida}}]{maeno2011evaluation}%
  \BibitemOpen
  \bibfield  {author} {\bibinfo {author} {\bibfnamefont {Y.}~\bibnamefont
  {Maeno}}, \bibinfo {author} {\bibfnamefont {S.}~\bibnamefont {Kittaka}},
  \bibinfo {author} {\bibfnamefont {T.}~\bibnamefont {Nomura}}, \bibinfo
  {author} {\bibfnamefont {S.}~\bibnamefont {Yonezawa}},\ and\ \bibinfo
  {author} {\bibfnamefont {K.}~\bibnamefont {Ishida}},\ }\bibfield  {title}
  {\bibinfo {title} {Evaluation of spin-triplet superconductivity in
  {Sr$_2$RuO$_4$}},\ }\href@noop {} {\bibfield  {journal} {\bibinfo  {journal}
  {J. Phys. Soc. Jpn.}\ }\textbf {\bibinfo {volume} {81}},\ \bibinfo {pages}
  {011009} (\bibinfo {year} {2011})}\BibitemShut {NoStop}%
\bibitem [{\citenamefont {Kallin}(2012)}]{kallin2012}%
  \BibitemOpen
  \bibfield  {author} {\bibinfo {author} {\bibfnamefont {C.}~\bibnamefont
  {Kallin}},\ }\bibfield  {title} {\bibinfo {title} {Chiral p-wave order in
  {Sr$_2$RuO$_4$}},\ }\href@noop {} {\bibfield  {journal} {\bibinfo  {journal}
  {Rep. Prog. Phys.}\ }\textbf {\bibinfo {volume} {75}},\ \bibinfo {pages}
  {042501} (\bibinfo {year} {2012})}\BibitemShut {NoStop}%
\bibitem [{\citenamefont {Liu}\ and\ \citenamefont
  {Mao}(2015)}]{liu2015unconventional}%
  \BibitemOpen
  \bibfield  {author} {\bibinfo {author} {\bibfnamefont {Y.}~\bibnamefont
  {Liu}}\ and\ \bibinfo {author} {\bibfnamefont {Z.-Q.}\ \bibnamefont {Mao}},\
  }\bibfield  {title} {\bibinfo {title} {Unconventional superconductivity in
  {Sr$_2$RuO$_4$}},\ }\href@noop {} {\bibfield  {journal} {\bibinfo  {journal}
  {Physica C}\ }\textbf {\bibinfo {volume} {514}},\ \bibinfo {pages} {339}
  (\bibinfo {year} {2015})}\BibitemShut {NoStop}%
\bibitem [{\citenamefont {Leggett}\ and\ \citenamefont
  {Liu}(2021)}]{leggett2021symmetry}%
  \BibitemOpen
  \bibfield  {author} {\bibinfo {author} {\bibfnamefont {A.~J.}\ \bibnamefont
  {Leggett}}\ and\ \bibinfo {author} {\bibfnamefont {Y.}~\bibnamefont {Liu}},\
  }\bibfield  {title} {\bibinfo {title} {Symmetry properties of superconducting
  order parameter in {Sr$_2$RuO$_4$}},\ }\href@noop {} {\bibfield  {journal}
  {\bibinfo  {journal} {J. Supercond. Novel Mag.}\ }\textbf {\bibinfo {volume}
  {34}},\ \bibinfo {pages} {1647} (\bibinfo {year} {2021})}\BibitemShut
  {NoStop}%
\bibitem [{\citenamefont {Bergemann}\ \emph {et~al.}(2003)\citenamefont
  {Bergemann}, \citenamefont {Mackenzie}, \citenamefont {Julian}, \citenamefont
  {Forsythe},\ and\ \citenamefont {Ohmichi}}]{bergemann2003quasi}%
  \BibitemOpen
  \bibfield  {author} {\bibinfo {author} {\bibfnamefont {C.}~\bibnamefont
  {Bergemann}}, \bibinfo {author} {\bibfnamefont {A.~P.}\ \bibnamefont
  {Mackenzie}}, \bibinfo {author} {\bibfnamefont {S.}~\bibnamefont {Julian}},
  \bibinfo {author} {\bibfnamefont {D.}~\bibnamefont {Forsythe}},\ and\
  \bibinfo {author} {\bibfnamefont {E.}~\bibnamefont {Ohmichi}},\ }\bibfield
  {title} {\bibinfo {title} {Quasi-two-dimensional {Fermi} liquid properties of
  the unconventional superconductor {Sr$_2$RuO$_4$}},\ }\href@noop {}
  {\bibfield  {journal} {\bibinfo  {journal} {Advances in Physics}\ }\textbf
  {\bibinfo {volume} {52}},\ \bibinfo {pages} {639} (\bibinfo {year}
  {2003})}\BibitemShut {NoStop}%
\bibitem [{\citenamefont {NishiZaki}\ \emph {et~al.}(2000)\citenamefont
  {NishiZaki}, \citenamefont {Maeno},\ and\ \citenamefont
  {Mao}}]{nishizaki2000changes}%
  \BibitemOpen
  \bibfield  {author} {\bibinfo {author} {\bibfnamefont {S.}~\bibnamefont
  {NishiZaki}}, \bibinfo {author} {\bibfnamefont {Y.}~\bibnamefont {Maeno}},\
  and\ \bibinfo {author} {\bibfnamefont {Z.}~\bibnamefont {Mao}},\ }\bibfield
  {title} {\bibinfo {title} {Changes in the superconducting state of
  {Sr$_2$RuO$_4$} under magnetic fields probed by specific heat},\ }\href@noop
  {} {\bibfield  {journal} {\bibinfo  {journal} {J. Phys. Soc. Jpn.}\ }\textbf
  {\bibinfo {volume} {69}},\ \bibinfo {pages} {572} (\bibinfo {year}
  {2000})}\BibitemShut {NoStop}%
\bibitem [{\citenamefont {Deguchi}\ \emph {et~al.}(2004)\citenamefont
  {Deguchi}, \citenamefont {Mao}, \citenamefont {Yaguchi},\ and\ \citenamefont
  {Maeno}}]{deguchi2004gap}%
  \BibitemOpen
  \bibfield  {author} {\bibinfo {author} {\bibfnamefont {K.}~\bibnamefont
  {Deguchi}}, \bibinfo {author} {\bibfnamefont {Z.}~\bibnamefont {Mao}},
  \bibinfo {author} {\bibfnamefont {H.}~\bibnamefont {Yaguchi}},\ and\ \bibinfo
  {author} {\bibfnamefont {Y.}~\bibnamefont {Maeno}},\ }\bibfield  {title}
  {\bibinfo {title} {Gap structure of the spin-triplet superconductor
  {Sr$_2$RuO$_4$} determined from the field-orientation dependence of the
  specific heat},\ }\href@noop {} {\bibfield  {journal} {\bibinfo  {journal}
  {Phys. Rev. Lett.}\ }\textbf {\bibinfo {volume} {92}},\ \bibinfo {pages}
  {047002} (\bibinfo {year} {2004})}\BibitemShut {NoStop}%
\bibitem [{\citenamefont {Kittaka}\ \emph {et~al.}(2018)\citenamefont
  {Kittaka}, \citenamefont {Nakamura}, \citenamefont {Sakakibara},
  \citenamefont {Kikugawa}, \citenamefont {Terashima}, \citenamefont {Uji},
  \citenamefont {Sokolov}, \citenamefont {Mackenzie}, \citenamefont {Irie},
  \citenamefont {Tsutsumi} \emph {et~al.}}]{kittaka2018searching}%
  \BibitemOpen
  \bibfield  {author} {\bibinfo {author} {\bibfnamefont {S.}~\bibnamefont
  {Kittaka}}, \bibinfo {author} {\bibfnamefont {S.}~\bibnamefont {Nakamura}},
  \bibinfo {author} {\bibfnamefont {T.}~\bibnamefont {Sakakibara}}, \bibinfo
  {author} {\bibfnamefont {N.}~\bibnamefont {Kikugawa}}, \bibinfo {author}
  {\bibfnamefont {T.}~\bibnamefont {Terashima}}, \bibinfo {author}
  {\bibfnamefont {S.}~\bibnamefont {Uji}}, \bibinfo {author} {\bibfnamefont
  {D.~A.}\ \bibnamefont {Sokolov}}, \bibinfo {author} {\bibfnamefont {A.~P.}\
  \bibnamefont {Mackenzie}}, \bibinfo {author} {\bibfnamefont {K.}~\bibnamefont
  {Irie}}, \bibinfo {author} {\bibfnamefont {Y.}~\bibnamefont {Tsutsumi}},
  \emph {et~al.},\ }\bibfield  {title} {\bibinfo {title} {Searching for gap
  zeros in {Sr$_2$RuO$_4$} via field-angle-dependent specific-heat
  measurement},\ }\href@noop {} {\bibfield  {journal} {\bibinfo  {journal} {J.
  Phys. Soc. Jpn.}\ }\textbf {\bibinfo {volume} {87}},\ \bibinfo {pages}
  {093703} (\bibinfo {year} {2018})}\BibitemShut {NoStop}%
\bibitem [{\citenamefont {Izawa}\ \emph {et~al.}(2001)\citenamefont {Izawa},
  \citenamefont {Takahashi}, \citenamefont {Yamaguchi}, \citenamefont
  {Matsuda}, \citenamefont {Suzuki}, \citenamefont {Sasaki}, \citenamefont
  {Fukase}, \citenamefont {Yoshida}, \citenamefont {Settai},\ and\
  \citenamefont {Onuki}}]{izawa2001thermal}%
  \BibitemOpen
  \bibfield  {author} {\bibinfo {author} {\bibfnamefont {K.}~\bibnamefont
  {Izawa}}, \bibinfo {author} {\bibfnamefont {H.}~\bibnamefont {Takahashi}},
  \bibinfo {author} {\bibfnamefont {H.}~\bibnamefont {Yamaguchi}}, \bibinfo
  {author} {\bibfnamefont {Y.}~\bibnamefont {Matsuda}}, \bibinfo {author}
  {\bibfnamefont {M.}~\bibnamefont {Suzuki}}, \bibinfo {author} {\bibfnamefont
  {T.}~\bibnamefont {Sasaki}}, \bibinfo {author} {\bibfnamefont
  {T.}~\bibnamefont {Fukase}}, \bibinfo {author} {\bibfnamefont
  {Y.}~\bibnamefont {Yoshida}}, \bibinfo {author} {\bibfnamefont
  {R.}~\bibnamefont {Settai}},\ and\ \bibinfo {author} {\bibfnamefont
  {Y.}~\bibnamefont {Onuki}},\ }\bibfield  {title} {\bibinfo {title}
  {Superconducting gap structure of spin-triplet superconductor {Sr$_2$RuO$_4$}
  studied by thermal conductivity},\ }\href@noop {} {\bibfield  {journal}
  {\bibinfo  {journal} {Phys. Rev. Lett.}\ }\textbf {\bibinfo {volume} {86}},\
  \bibinfo {pages} {2653} (\bibinfo {year} {2001})}\BibitemShut {NoStop}%
\bibitem [{\citenamefont {Suzuki}\ \emph {et~al.}(2002)\citenamefont {Suzuki},
  \citenamefont {Tanatar}, \citenamefont {Kikugawa}, \citenamefont {Mao},
  \citenamefont {Maeno},\ and\ \citenamefont {Ishiguro}}]{suzuki2002universal}%
  \BibitemOpen
  \bibfield  {author} {\bibinfo {author} {\bibfnamefont {M.}~\bibnamefont
  {Suzuki}}, \bibinfo {author} {\bibfnamefont {M.}~\bibnamefont {Tanatar}},
  \bibinfo {author} {\bibfnamefont {N.}~\bibnamefont {Kikugawa}}, \bibinfo
  {author} {\bibfnamefont {Z.}~\bibnamefont {Mao}}, \bibinfo {author}
  {\bibfnamefont {Y.}~\bibnamefont {Maeno}},\ and\ \bibinfo {author}
  {\bibfnamefont {T.}~\bibnamefont {Ishiguro}},\ }\bibfield  {title} {\bibinfo
  {title} {Universal heat transport in {Sr$_2$RuO$_4$}},\ }\href@noop {}
  {\bibfield  {journal} {\bibinfo  {journal} {Phys. Rev. Lett.}\ }\textbf
  {\bibinfo {volume} {88}},\ \bibinfo {pages} {227004} (\bibinfo {year}
  {2002})}\BibitemShut {NoStop}%
\bibitem [{\citenamefont {Hassinger}\ \emph {et~al.}(2017)\citenamefont
  {Hassinger}, \citenamefont {Bourgeois-Hope}, \citenamefont {Taniguchi},
  \citenamefont {de~Cotret}, \citenamefont {Grissonnanche}, \citenamefont
  {Anwar}, \citenamefont {Maeno}, \citenamefont {Doiron-Leyraud},\ and\
  \citenamefont {Taillefer}}]{hassinger2017vertical}%
  \BibitemOpen
  \bibfield  {author} {\bibinfo {author} {\bibfnamefont {E.}~\bibnamefont
  {Hassinger}}, \bibinfo {author} {\bibfnamefont {P.}~\bibnamefont
  {Bourgeois-Hope}}, \bibinfo {author} {\bibfnamefont {H.}~\bibnamefont
  {Taniguchi}}, \bibinfo {author} {\bibfnamefont {S.~R.}\ \bibnamefont
  {de~Cotret}}, \bibinfo {author} {\bibfnamefont {G.}~\bibnamefont
  {Grissonnanche}}, \bibinfo {author} {\bibfnamefont {M.~S.}\ \bibnamefont
  {Anwar}}, \bibinfo {author} {\bibfnamefont {Y.}~\bibnamefont {Maeno}},
  \bibinfo {author} {\bibfnamefont {N.}~\bibnamefont {Doiron-Leyraud}},\ and\
  \bibinfo {author} {\bibfnamefont {L.}~\bibnamefont {Taillefer}},\ }\bibfield
  {title} {\bibinfo {title} {Vertical line nodes in the superconducting gap
  structure of {Sr$_2$RuO$_4$}},\ }\href@noop {} {\bibfield  {journal}
  {\bibinfo  {journal} {Phys. Rev. X}\ }\textbf {\bibinfo {volume} {7}},\
  \bibinfo {pages} {011032} (\bibinfo {year} {2017})}\BibitemShut {NoStop}%
\bibitem [{\citenamefont {Hirschfeld}\ and\ \citenamefont
  {Goldenfeld}(1993)}]{hirschfeld1993effect}%
  \BibitemOpen
  \bibfield  {author} {\bibinfo {author} {\bibfnamefont {P.~J.}\ \bibnamefont
  {Hirschfeld}}\ and\ \bibinfo {author} {\bibfnamefont {N.}~\bibnamefont
  {Goldenfeld}},\ }\bibfield  {title} {\bibinfo {title} {Effect of strong
  scattering on the low-temperature penetration depth of a {$d$}-wave
  superconductor},\ }\href@noop {} {\bibfield  {journal} {\bibinfo  {journal}
  {Phys. Rev. B}\ }\textbf {\bibinfo {volume} {48}},\ \bibinfo {pages} {4219}
  (\bibinfo {year} {1993})}\BibitemShut {NoStop}%
\bibitem [{\citenamefont {Bonalde}\ \emph {et~al.}(2000)\citenamefont
  {Bonalde}, \citenamefont {Yanoff}, \citenamefont {Salamon}, \citenamefont
  {Van~Harlingen}, \citenamefont {Chia}, \citenamefont {Mao},\ and\
  \citenamefont {Maeno}}]{bonalde2000temperature}%
  \BibitemOpen
  \bibfield  {author} {\bibinfo {author} {\bibfnamefont {I.}~\bibnamefont
  {Bonalde}}, \bibinfo {author} {\bibfnamefont {B.~D.}\ \bibnamefont {Yanoff}},
  \bibinfo {author} {\bibfnamefont {M.}~\bibnamefont {Salamon}}, \bibinfo
  {author} {\bibfnamefont {D.}~\bibnamefont {Van~Harlingen}}, \bibinfo {author}
  {\bibfnamefont {E.}~\bibnamefont {Chia}}, \bibinfo {author} {\bibfnamefont
  {Z.}~\bibnamefont {Mao}},\ and\ \bibinfo {author} {\bibfnamefont
  {Y.}~\bibnamefont {Maeno}},\ }\bibfield  {title} {\bibinfo {title}
  {Temperature dependence of the penetration depth in {Sr$_2$RuO$_4$}: evidence
  for nodes in the gap function},\ }\href@noop {} {\bibfield  {journal}
  {\bibinfo  {journal} {Phys. Rev. Lett.}\ }\textbf {\bibinfo {volume} {85}},\
  \bibinfo {pages} {4775} (\bibinfo {year} {2000})}\BibitemShut {NoStop}%
\bibitem [{\citenamefont {Ormeno}\ \emph {et~al.}(2006)\citenamefont {Ormeno},
  \citenamefont {Hein}, \citenamefont {Barraclough}, \citenamefont {Sibley},
  \citenamefont {Gough}, \citenamefont {Mao}, \citenamefont {Nishizaki},\ and\
  \citenamefont {Maeno}}]{ormeno2006electrodynamic}%
  \BibitemOpen
  \bibfield  {author} {\bibinfo {author} {\bibfnamefont {R.}~\bibnamefont
  {Ormeno}}, \bibinfo {author} {\bibfnamefont {M.}~\bibnamefont {Hein}},
  \bibinfo {author} {\bibfnamefont {T.}~\bibnamefont {Barraclough}}, \bibinfo
  {author} {\bibfnamefont {A.}~\bibnamefont {Sibley}}, \bibinfo {author}
  {\bibfnamefont {C.}~\bibnamefont {Gough}}, \bibinfo {author} {\bibfnamefont
  {Z.}~\bibnamefont {Mao}}, \bibinfo {author} {\bibfnamefont {S.}~\bibnamefont
  {Nishizaki}},\ and\ \bibinfo {author} {\bibfnamefont {Y.}~\bibnamefont
  {Maeno}},\ }\bibfield  {title} {\bibinfo {title} {Electrodynamic response of
  {Sr$_2$RuO$_4$}},\ }\href@noop {} {\bibfield  {journal} {\bibinfo  {journal}
  {Phys. Rev. B}\ }\textbf {\bibinfo {volume} {74}},\ \bibinfo {pages} {092504}
  (\bibinfo {year} {2006})}\BibitemShut {NoStop}%
\bibitem [{\citenamefont {Baker}\ \emph {et~al.}(2009)\citenamefont {Baker},
  \citenamefont {Ormeno}, \citenamefont {Gough}, \citenamefont {Mao},
  \citenamefont {Nishizaki},\ and\ \citenamefont {Maeno}}]{baker2009microwave}%
  \BibitemOpen
  \bibfield  {author} {\bibinfo {author} {\bibfnamefont {P.}~\bibnamefont
  {Baker}}, \bibinfo {author} {\bibfnamefont {R.}~\bibnamefont {Ormeno}},
  \bibinfo {author} {\bibfnamefont {C.}~\bibnamefont {Gough}}, \bibinfo
  {author} {\bibfnamefont {Z.}~\bibnamefont {Mao}}, \bibinfo {author}
  {\bibfnamefont {S.}~\bibnamefont {Nishizaki}},\ and\ \bibinfo {author}
  {\bibfnamefont {Y.}~\bibnamefont {Maeno}},\ }\bibfield  {title} {\bibinfo
  {title} {Microwave surface impedance measurements of {Sr$_2$RuO$_4$}: The
  effect of impurities},\ }\href@noop {} {\bibfield  {journal} {\bibinfo
  {journal} {Phys. Rev. B}\ }\textbf {\bibinfo {volume} {80}},\ \bibinfo
  {pages} {115126} (\bibinfo {year} {2009})}\BibitemShut {NoStop}%
\bibitem [{\citenamefont {Mao}\ \emph {et~al.}(2000)\citenamefont {Mao},
  \citenamefont {Maeno},\ and\ \citenamefont {Fukazawa}}]{mao2000crystal}%
  \BibitemOpen
  \bibfield  {author} {\bibinfo {author} {\bibfnamefont {Z.}~\bibnamefont
  {Mao}}, \bibinfo {author} {\bibfnamefont {Y.}~\bibnamefont {Maeno}},\ and\
  \bibinfo {author} {\bibfnamefont {H.}~\bibnamefont {Fukazawa}},\ }\bibfield
  {title} {\bibinfo {title} {Crystal growth of {Sr$_2$RuO$_4$}},\ }\href@noop
  {} {\bibfield  {journal} {\bibinfo  {journal} {Materials Research Bulletin}\
  }\textbf {\bibinfo {volume} {35}},\ \bibinfo {pages} {1813} (\bibinfo {year}
  {2000})}\BibitemShut {NoStop}%
\bibitem [{\citenamefont {Kosztin}\ and\ \citenamefont
  {Leggett}(1997)}]{kosztin1997nonlocal}%
  \BibitemOpen
  \bibfield  {author} {\bibinfo {author} {\bibfnamefont {I.}~\bibnamefont
  {Kosztin}}\ and\ \bibinfo {author} {\bibfnamefont {A.~J.}\ \bibnamefont
  {Leggett}},\ }\bibfield  {title} {\bibinfo {title} {Nonlocal effects on the
  magnetic penetration depth in d-wave superconductors},\ }\href@noop {}
  {\bibfield  {journal} {\bibinfo  {journal} {Phys. Rev. Lett.}\ }\textbf
  {\bibinfo {volume} {79}},\ \bibinfo {pages} {135} (\bibinfo {year}
  {1997})}\BibitemShut {NoStop}%
\bibitem [{\citenamefont {Steppke}\ \emph {et~al.}(2017)\citenamefont
  {Steppke}, \citenamefont {Zhao}, \citenamefont {Barber}, \citenamefont
  {Scaffidi}, \citenamefont {Jerzembeck}, \citenamefont {Rosner}, \citenamefont
  {Gibbs}, \citenamefont {Maeno}, \citenamefont {Simon}, \citenamefont
  {Mackenzie},\ and\ \citenamefont {Hicks}}]{steppke2017strong}%
  \BibitemOpen
  \bibfield  {author} {\bibinfo {author} {\bibfnamefont {A.}~\bibnamefont
  {Steppke}}, \bibinfo {author} {\bibfnamefont {L.}~\bibnamefont {Zhao}},
  \bibinfo {author} {\bibfnamefont {M.~E.}\ \bibnamefont {Barber}}, \bibinfo
  {author} {\bibfnamefont {T.}~\bibnamefont {Scaffidi}}, \bibinfo {author}
  {\bibfnamefont {F.}~\bibnamefont {Jerzembeck}}, \bibinfo {author}
  {\bibfnamefont {H.}~\bibnamefont {Rosner}}, \bibinfo {author} {\bibfnamefont
  {A.~S.}\ \bibnamefont {Gibbs}}, \bibinfo {author} {\bibfnamefont
  {Y.}~\bibnamefont {Maeno}}, \bibinfo {author} {\bibfnamefont {S.~H.}\
  \bibnamefont {Simon}}, \bibinfo {author} {\bibfnamefont {A.~P.}\ \bibnamefont
  {Mackenzie}},\ and\ \bibinfo {author} {\bibfnamefont {C.~W.}\ \bibnamefont
  {Hicks}},\ }\bibfield  {title} {\bibinfo {title} {Strong peak in {$T_c$} of
  {Sr$_2$RuO$_4$} under uniaxial pressure},\ }\href@noop {} {\bibfield
  {journal} {\bibinfo  {journal} {Science}\ }\textbf {\bibinfo {volume}
  {355}},\ \bibinfo {pages} {eaaf9398} (\bibinfo {year} {2017})}\BibitemShut
  {NoStop}%
\bibitem [{\citenamefont {Sunko}\ \emph {et~al.}(2019)\citenamefont {Sunko},
  \citenamefont {Abarca~Morales}, \citenamefont {Markovi{\'c}}, \citenamefont
  {Barber}, \citenamefont {Milosavljevi{\'c}}, \citenamefont {Mazzola},
  \citenamefont {Sokolov}, \citenamefont {Kikugawa}, \citenamefont {Cacho},
  \citenamefont {Dudin} \emph {et~al.}}]{sunko2019direct}%
  \BibitemOpen
  \bibfield  {author} {\bibinfo {author} {\bibfnamefont {V.}~\bibnamefont
  {Sunko}}, \bibinfo {author} {\bibfnamefont {E.}~\bibnamefont
  {Abarca~Morales}}, \bibinfo {author} {\bibfnamefont {I.}~\bibnamefont
  {Markovi{\'c}}}, \bibinfo {author} {\bibfnamefont {M.~E.}\ \bibnamefont
  {Barber}}, \bibinfo {author} {\bibfnamefont {D.}~\bibnamefont
  {Milosavljevi{\'c}}}, \bibinfo {author} {\bibfnamefont {F.}~\bibnamefont
  {Mazzola}}, \bibinfo {author} {\bibfnamefont {D.~A.}\ \bibnamefont
  {Sokolov}}, \bibinfo {author} {\bibfnamefont {N.}~\bibnamefont {Kikugawa}},
  \bibinfo {author} {\bibfnamefont {C.}~\bibnamefont {Cacho}}, \bibinfo
  {author} {\bibfnamefont {P.}~\bibnamefont {Dudin}}, \emph {et~al.},\
  }\bibfield  {title} {\bibinfo {title} {Direct observation of a uniaxial
  stress-driven {Lifshitz} transition in {Sr$_2$RuO$_4$}},\ }\href@noop {}
  {\bibfield  {journal} {\bibinfo  {journal} {npj Quantum Mater.}\ }\textbf
  {\bibinfo {volume} {4}},\ \bibinfo {pages} {46} (\bibinfo {year}
  {2019})}\BibitemShut {NoStop}%
\bibitem [{\citenamefont {Hicks}\ \emph
  {et~al.}(2014{\natexlab{a}})\citenamefont {Hicks}, \citenamefont {Brodsky},
  \citenamefont {Yelland}, \citenamefont {Gibbs}, \citenamefont {Bruin},
  \citenamefont {Barber}, \citenamefont {Edkins}, \citenamefont {Nishimura},
  \citenamefont {Yonezawa}, \citenamefont {Maeno}, \citenamefont {Mackenzie},
  \citenamefont {Baghi}, \citenamefont {Hicks}, \citenamefont {Brodsky},\ and\
  \citenamefont {Edward}}]{hicks2014strong}%
  \BibitemOpen
  \bibfield  {author} {\bibinfo {author} {\bibfnamefont {C.~W.}\ \bibnamefont
  {Hicks}}, \bibinfo {author} {\bibfnamefont {D.~O.}\ \bibnamefont {Brodsky}},
  \bibinfo {author} {\bibfnamefont {E.~A.}\ \bibnamefont {Yelland}}, \bibinfo
  {author} {\bibfnamefont {A.~S.}\ \bibnamefont {Gibbs}}, \bibinfo {author}
  {\bibfnamefont {J.~A.~N.}\ \bibnamefont {Bruin}}, \bibinfo {author}
  {\bibfnamefont {M.~E.}\ \bibnamefont {Barber}}, \bibinfo {author}
  {\bibfnamefont {S.~D.}\ \bibnamefont {Edkins}}, \bibinfo {author}
  {\bibfnamefont {K.}~\bibnamefont {Nishimura}}, \bibinfo {author}
  {\bibfnamefont {S.}~\bibnamefont {Yonezawa}}, \bibinfo {author}
  {\bibfnamefont {Y.}~\bibnamefont {Maeno}}, \bibinfo {author} {\bibfnamefont
  {A.~P.}\ \bibnamefont {Mackenzie}}, \bibinfo {author} {\bibfnamefont
  {B.}~\bibnamefont {Baghi}}, \bibinfo {author} {\bibfnamefont {C.~W.}\
  \bibnamefont {Hicks}}, \bibinfo {author} {\bibfnamefont {D.}~\bibnamefont
  {Brodsky}},\ and\ \bibinfo {author} {\bibfnamefont {A.}~\bibnamefont
  {Edward}},\ }\bibfield  {title} {\bibinfo {title} {Strong increase of {$T_c$}
  of {Sr$_2$RuO$_4$} under both tensile and compressive strain},\ }\href@noop
  {} {\bibfield  {journal} {\bibinfo  {journal} {Science}\ }\textbf {\bibinfo
  {volume} {344}},\ \bibinfo {pages} {283} (\bibinfo {year}
  {2014}{\natexlab{a}})}\BibitemShut {NoStop}%
\bibitem [{\citenamefont {Jerzembeck}\ \emph {et~al.}(2023)\citenamefont
  {Jerzembeck}, \citenamefont {Steppke}, \citenamefont {Pustogow},
  \citenamefont {Luo}, \citenamefont {Chronister}, \citenamefont {Sokolov},
  \citenamefont {Kikugawa}, \citenamefont {Li}, \citenamefont {Nicklas},
  \citenamefont {Brown} \emph {et~al.}}]{jerzembeck2023upper}%
  \BibitemOpen
  \bibfield  {author} {\bibinfo {author} {\bibfnamefont {F.}~\bibnamefont
  {Jerzembeck}}, \bibinfo {author} {\bibfnamefont {A.}~\bibnamefont {Steppke}},
  \bibinfo {author} {\bibfnamefont {A.}~\bibnamefont {Pustogow}}, \bibinfo
  {author} {\bibfnamefont {Y.}~\bibnamefont {Luo}}, \bibinfo {author}
  {\bibfnamefont {A.}~\bibnamefont {Chronister}}, \bibinfo {author}
  {\bibfnamefont {D.~A.}\ \bibnamefont {Sokolov}}, \bibinfo {author}
  {\bibfnamefont {N.}~\bibnamefont {Kikugawa}}, \bibinfo {author}
  {\bibfnamefont {Y.-S.}\ \bibnamefont {Li}}, \bibinfo {author} {\bibfnamefont
  {M.}~\bibnamefont {Nicklas}}, \bibinfo {author} {\bibfnamefont {S.~E.}\
  \bibnamefont {Brown}}, \emph {et~al.},\ }\bibfield  {title} {\bibinfo {title}
  {Upper critical field of {Sr$_2$RuO$_4$} under in-plane uniaxial pressure},\
  }\href@noop {} {\bibfield  {journal} {\bibinfo  {journal} {Phys. Rev. B}\
  }\textbf {\bibinfo {volume} {107}},\ \bibinfo {pages} {064509} (\bibinfo
  {year} {2023})}\BibitemShut {NoStop}%
\bibitem [{\citenamefont {Hsu}\ \emph {et~al.}(2016)\citenamefont {Hsu},
  \citenamefont {Cho}, \citenamefont {Rebola}, \citenamefont {Burganov},
  \citenamefont {Adamo}, \citenamefont {Shen}, \citenamefont {Schlom},
  \citenamefont {Fennie},\ and\ \citenamefont {Kim}}]{hsu2016manipulating}%
  \BibitemOpen
  \bibfield  {author} {\bibinfo {author} {\bibfnamefont {Y.-T.}\ \bibnamefont
  {Hsu}}, \bibinfo {author} {\bibfnamefont {W.}~\bibnamefont {Cho}}, \bibinfo
  {author} {\bibfnamefont {A.~F.}\ \bibnamefont {Rebola}}, \bibinfo {author}
  {\bibfnamefont {B.}~\bibnamefont {Burganov}}, \bibinfo {author}
  {\bibfnamefont {C.}~\bibnamefont {Adamo}}, \bibinfo {author} {\bibfnamefont
  {K.~M.}\ \bibnamefont {Shen}}, \bibinfo {author} {\bibfnamefont {D.~G.}\
  \bibnamefont {Schlom}}, \bibinfo {author} {\bibfnamefont {C.~J.}\
  \bibnamefont {Fennie}},\ and\ \bibinfo {author} {\bibfnamefont {E.-A.}\
  \bibnamefont {Kim}},\ }\bibfield  {title} {\bibinfo {title} {Manipulating
  superconductivity in ruthenates through {Fermi} surface engineering},\
  }\href@noop {} {\bibfield  {journal} {\bibinfo  {journal} {Phys. Rev. B}\
  }\textbf {\bibinfo {volume} {94}},\ \bibinfo {pages} {045118} (\bibinfo
  {year} {2016})}\BibitemShut {NoStop}%
\bibitem [{\citenamefont {Mravlje}\ \emph {et~al.}(2011)\citenamefont
  {Mravlje}, \citenamefont {Aichhorn}, \citenamefont {Miyake}, \citenamefont
  {Haule}, \citenamefont {Kotliar},\ and\ \citenamefont
  {Georges}}]{mravlje2011coherence}%
  \BibitemOpen
  \bibfield  {author} {\bibinfo {author} {\bibfnamefont {J.}~\bibnamefont
  {Mravlje}}, \bibinfo {author} {\bibfnamefont {M.}~\bibnamefont {Aichhorn}},
  \bibinfo {author} {\bibfnamefont {T.}~\bibnamefont {Miyake}}, \bibinfo
  {author} {\bibfnamefont {K.}~\bibnamefont {Haule}}, \bibinfo {author}
  {\bibfnamefont {G.}~\bibnamefont {Kotliar}},\ and\ \bibinfo {author}
  {\bibfnamefont {A.}~\bibnamefont {Georges}},\ }\bibfield  {title} {\bibinfo
  {title} {Coherence-incoherence crossover and the mass-renormalization puzzles
  in {Sr$_2$RuO$_4$}},\ }\href@noop {} {\bibfield  {journal} {\bibinfo
  {journal} {Phys. Rev. Lett.}\ }\textbf {\bibinfo {volume} {106}},\ \bibinfo
  {pages} {096401} (\bibinfo {year} {2011})}\BibitemShut {NoStop}%
\bibitem [{\citenamefont {Mueller}\ \emph {et~al.}(2023)\citenamefont
  {Mueller}, \citenamefont {Iguchi}, \citenamefont {Watson}, \citenamefont
  {Hicks}, \citenamefont {Maeno},\ and\ \citenamefont
  {Moler}}]{mueller2023constraints}%
  \BibitemOpen
  \bibfield  {author} {\bibinfo {author} {\bibfnamefont {E.}~\bibnamefont
  {Mueller}}, \bibinfo {author} {\bibfnamefont {Y.}~\bibnamefont {Iguchi}},
  \bibinfo {author} {\bibfnamefont {C.}~\bibnamefont {Watson}}, \bibinfo
  {author} {\bibfnamefont {C.~W.}\ \bibnamefont {Hicks}}, \bibinfo {author}
  {\bibfnamefont {Y.}~\bibnamefont {Maeno}},\ and\ \bibinfo {author}
  {\bibfnamefont {K.~A.}\ \bibnamefont {Moler}},\ }\bibfield  {title} {\bibinfo
  {title} {Constraints on a split superconducting transition under uniaxial
  strain in {Sr$_2$RuO$_4$} from scanning squid microscopy},\ }\href@noop {}
  {\bibfield  {journal} {\bibinfo  {journal} {Phys. Rev. B}\ }\textbf {\bibinfo
  {volume} {108}},\ \bibinfo {pages} {144501} (\bibinfo {year}
  {2023})}\BibitemShut {NoStop}%
\bibitem [{\citenamefont {Grinenko}\ \emph {et~al.}(2021)\citenamefont
  {Grinenko}, \citenamefont {Ghosh}, \citenamefont {Sarkar}, \citenamefont
  {Orain}, \citenamefont {Nikitin}, \citenamefont {Elender}, \citenamefont
  {Das}, \citenamefont {Guguchia}, \citenamefont {Br{\"u}ckner}, \citenamefont
  {Barber}, \citenamefont {Park}, \citenamefont {Kikugawa}, \citenamefont
  {Sokolov}, \citenamefont {Bobowski}, \citenamefont {Miyoshi}, \citenamefont
  {Maeno}, \citenamefont {Mackenzie}, \citenamefont {Luetkens}, \citenamefont
  {Hicks},\ and\ \citenamefont {Klauss}}]{grinenko2021split}%
  \BibitemOpen
  \bibfield  {author} {\bibinfo {author} {\bibfnamefont {V.}~\bibnamefont
  {Grinenko}}, \bibinfo {author} {\bibfnamefont {S.}~\bibnamefont {Ghosh}},
  \bibinfo {author} {\bibfnamefont {R.}~\bibnamefont {Sarkar}}, \bibinfo
  {author} {\bibfnamefont {J.~C.}\ \bibnamefont {Orain}}, \bibinfo {author}
  {\bibfnamefont {A.}~\bibnamefont {Nikitin}}, \bibinfo {author} {\bibfnamefont
  {M.}~\bibnamefont {Elender}}, \bibinfo {author} {\bibfnamefont
  {D.}~\bibnamefont {Das}}, \bibinfo {author} {\bibfnamefont {Z.}~\bibnamefont
  {Guguchia}}, \bibinfo {author} {\bibfnamefont {F.}~\bibnamefont
  {Br{\"u}ckner}}, \bibinfo {author} {\bibfnamefont {M.~E.}\ \bibnamefont
  {Barber}}, \bibinfo {author} {\bibfnamefont {J.}~\bibnamefont {Park}},
  \bibinfo {author} {\bibfnamefont {N.}~\bibnamefont {Kikugawa}}, \bibinfo
  {author} {\bibfnamefont {D.~A.}\ \bibnamefont {Sokolov}}, \bibinfo {author}
  {\bibfnamefont {J.~S.}\ \bibnamefont {Bobowski}}, \bibinfo {author}
  {\bibfnamefont {T.}~\bibnamefont {Miyoshi}}, \bibinfo {author} {\bibfnamefont
  {Y.}~\bibnamefont {Maeno}}, \bibinfo {author} {\bibfnamefont {A.~P.}\
  \bibnamefont {Mackenzie}}, \bibinfo {author} {\bibfnamefont {H.}~\bibnamefont
  {Luetkens}}, \bibinfo {author} {\bibfnamefont {C.~W.}\ \bibnamefont
  {Hicks}},\ and\ \bibinfo {author} {\bibfnamefont {H.~H.}\ \bibnamefont
  {Klauss}},\ }\bibfield  {title} {\bibinfo {title} {Split superconducting and
  time-reversal symmetry-breaking transitions in {Sr$_2$RuO$_4$} under
  stress},\ }\href@noop {} {\bibfield  {journal} {\bibinfo  {journal} {Nat.
  Phys.}\ }\textbf {\bibinfo {volume} {17}},\ \bibinfo {pages} {748} (\bibinfo
  {year} {2021})}\BibitemShut {NoStop}%
\bibitem [{\citenamefont {Kirtley}\ \emph {et~al.}(2016)\citenamefont
  {Kirtley}, \citenamefont {Paulius}, \citenamefont {Rosenberg}, \citenamefont
  {Palmstrom}, \citenamefont {Holland}, \citenamefont {Spanton}, \citenamefont
  {Schiessl}, \citenamefont {Jermain}, \citenamefont {Gibbons}, \citenamefont
  {Fung}, \citenamefont {Huber}, \citenamefont {Ralph}, \citenamefont
  {Ketchen}, \citenamefont {Gibson},\ and\ \citenamefont
  {Moler}}]{kirtley2016scanning}%
  \BibitemOpen
  \bibfield  {author} {\bibinfo {author} {\bibfnamefont {J.~R.}\ \bibnamefont
  {Kirtley}}, \bibinfo {author} {\bibfnamefont {L.}~\bibnamefont {Paulius}},
  \bibinfo {author} {\bibfnamefont {A.~J.}\ \bibnamefont {Rosenberg}}, \bibinfo
  {author} {\bibfnamefont {J.~C.}\ \bibnamefont {Palmstrom}}, \bibinfo {author}
  {\bibfnamefont {C.~M.}\ \bibnamefont {Holland}}, \bibinfo {author}
  {\bibfnamefont {E.~M.}\ \bibnamefont {Spanton}}, \bibinfo {author}
  {\bibfnamefont {D.}~\bibnamefont {Schiessl}}, \bibinfo {author}
  {\bibfnamefont {C.~L.}\ \bibnamefont {Jermain}}, \bibinfo {author}
  {\bibfnamefont {J.}~\bibnamefont {Gibbons}}, \bibinfo {author} {\bibfnamefont
  {Y.~K.}\ \bibnamefont {Fung}}, \bibinfo {author} {\bibfnamefont {M.~E.}\
  \bibnamefont {Huber}}, \bibinfo {author} {\bibfnamefont {D.~C.}\ \bibnamefont
  {Ralph}}, \bibinfo {author} {\bibfnamefont {M.~B.}\ \bibnamefont {Ketchen}},
  \bibinfo {author} {\bibfnamefont {G.~W.}\ \bibnamefont {Gibson}},\ and\
  \bibinfo {author} {\bibfnamefont {K.~A.}\ \bibnamefont {Moler}},\ }\bibfield
  {title} {\bibinfo {title} {Scanning {SQUID} susceptometers with sub-micron
  spatial resolution},\ }\href@noop {} {\bibfield  {journal} {\bibinfo
  {journal} {Rev. Sci. Instrum.}\ }\textbf {\bibinfo {volume} {87}},\ \bibinfo
  {pages} {093702} (\bibinfo {year} {2016})}\BibitemShut {NoStop}%
\bibitem [{\citenamefont {Kirtley}\ \emph {et~al.}(2012)\citenamefont
  {Kirtley}, \citenamefont {Kalisky}, \citenamefont {Bert}, \citenamefont
  {Bell}, \citenamefont {Kim}, \citenamefont {Hikita}, \citenamefont {Hwang},
  \citenamefont {Ngai}, \citenamefont {Segal}, \citenamefont {Walker},
  \citenamefont {Ahn},\ and\ \citenamefont {Moler}}]{kirtley2012scanning}%
  \BibitemOpen
  \bibfield  {author} {\bibinfo {author} {\bibfnamefont {J.~R.}\ \bibnamefont
  {Kirtley}}, \bibinfo {author} {\bibfnamefont {B.}~\bibnamefont {Kalisky}},
  \bibinfo {author} {\bibfnamefont {J.~A.}\ \bibnamefont {Bert}}, \bibinfo
  {author} {\bibfnamefont {C.}~\bibnamefont {Bell}}, \bibinfo {author}
  {\bibfnamefont {M.}~\bibnamefont {Kim}}, \bibinfo {author} {\bibfnamefont
  {Y.}~\bibnamefont {Hikita}}, \bibinfo {author} {\bibfnamefont {H.~Y.}\
  \bibnamefont {Hwang}}, \bibinfo {author} {\bibfnamefont {J.~H.}\ \bibnamefont
  {Ngai}}, \bibinfo {author} {\bibfnamefont {Y.}~\bibnamefont {Segal}},
  \bibinfo {author} {\bibfnamefont {F.~J.}\ \bibnamefont {Walker}}, \bibinfo
  {author} {\bibfnamefont {C.~H.}\ \bibnamefont {Ahn}},\ and\ \bibinfo {author}
  {\bibfnamefont {K.~A.}\ \bibnamefont {Moler}},\ }\bibfield  {title} {\bibinfo
  {title} {Scanning {SQUID} susceptometry of a paramagnetic superconductor},\
  }\href@noop {} {\bibfield  {journal} {\bibinfo  {journal} {Phys. Rev. B}\
  }\textbf {\bibinfo {volume} {85}},\ \bibinfo {pages} {224518} (\bibinfo
  {year} {2012})}\BibitemShut {NoStop}%
\bibitem [{\citenamefont {Hicks}\ \emph
  {et~al.}(2014{\natexlab{b}})\citenamefont {Hicks}, \citenamefont {Barber},
  \citenamefont {Edkins}, \citenamefont {Brodsky},\ and\ \citenamefont
  {Mackenzie}}]{hicks2014piezoelectric}%
  \BibitemOpen
  \bibfield  {author} {\bibinfo {author} {\bibfnamefont {C.~W.}\ \bibnamefont
  {Hicks}}, \bibinfo {author} {\bibfnamefont {M.~E.}\ \bibnamefont {Barber}},
  \bibinfo {author} {\bibfnamefont {S.~D.}\ \bibnamefont {Edkins}}, \bibinfo
  {author} {\bibfnamefont {D.~O.}\ \bibnamefont {Brodsky}},\ and\ \bibinfo
  {author} {\bibfnamefont {A.~P.}\ \bibnamefont {Mackenzie}},\ }\bibfield
  {title} {\bibinfo {title} {Piezoelectric-based apparatus for strain tuning},\
  }\href@noop {} {\bibfield  {journal} {\bibinfo  {journal} {Rev. Sci.
  Instrum.}\ }\textbf {\bibinfo {volume} {85}},\ \bibinfo {pages} {065003}
  (\bibinfo {year} {2014}{\natexlab{b}})}\BibitemShut {NoStop}%
\bibitem [{\citenamefont {Watson}\ \emph {et~al.}(2018)\citenamefont {Watson},
  \citenamefont {Gibbs}, \citenamefont {Mackenzie}, \citenamefont {Hicks},\
  and\ \citenamefont {Moler}}]{watson2018micron}%
  \BibitemOpen
  \bibfield  {author} {\bibinfo {author} {\bibfnamefont {C.~A.}\ \bibnamefont
  {Watson}}, \bibinfo {author} {\bibfnamefont {A.~S.}\ \bibnamefont {Gibbs}},
  \bibinfo {author} {\bibfnamefont {A.~P.}\ \bibnamefont {Mackenzie}}, \bibinfo
  {author} {\bibfnamefont {C.~W.}\ \bibnamefont {Hicks}},\ and\ \bibinfo
  {author} {\bibfnamefont {K.~A.}\ \bibnamefont {Moler}},\ }\bibfield  {title}
  {\bibinfo {title} {Micron-scale measurements of low anisotropic strain
  response of local {$T_c$} in {Sr$_2$RuO$_4$}},\ }\href@noop {} {\bibfield
  {journal} {\bibinfo  {journal} {Phys. Rev. B}\ }\textbf {\bibinfo {volume}
  {98}},\ \bibinfo {pages} {094521} (\bibinfo {year} {2018})}\BibitemShut
  {NoStop}%
\bibitem [{\citenamefont {Barber}\ \emph {et~al.}(2019)\citenamefont {Barber},
  \citenamefont {Lechermann}, \citenamefont {Streltsov}, \citenamefont
  {Skornyakov}, \citenamefont {Ghosh}, \citenamefont {Ramshaw}, \citenamefont
  {Kikugawa}, \citenamefont {Sokolov}, \citenamefont {Mackenzie}, \citenamefont
  {Hicks} \emph {et~al.}}]{barber2019role}%
  \BibitemOpen
  \bibfield  {author} {\bibinfo {author} {\bibfnamefont {M.~E.}\ \bibnamefont
  {Barber}}, \bibinfo {author} {\bibfnamefont {F.}~\bibnamefont {Lechermann}},
  \bibinfo {author} {\bibfnamefont {S.~V.}\ \bibnamefont {Streltsov}}, \bibinfo
  {author} {\bibfnamefont {S.~L.}\ \bibnamefont {Skornyakov}}, \bibinfo
  {author} {\bibfnamefont {S.}~\bibnamefont {Ghosh}}, \bibinfo {author}
  {\bibfnamefont {B.}~\bibnamefont {Ramshaw}}, \bibinfo {author} {\bibfnamefont
  {N.}~\bibnamefont {Kikugawa}}, \bibinfo {author} {\bibfnamefont {D.~A.}\
  \bibnamefont {Sokolov}}, \bibinfo {author} {\bibfnamefont {A.~P.}\
  \bibnamefont {Mackenzie}}, \bibinfo {author} {\bibfnamefont {C.~W.}\
  \bibnamefont {Hicks}}, \emph {et~al.},\ }\bibfield  {title} {\bibinfo {title}
  {Role of correlations in determining the {Van Hove} strain in
  {Sr$_2$RuO$_4$}},\ }\href@noop {} {\bibfield  {journal} {\bibinfo  {journal}
  {Phys. Rev. B}\ }\textbf {\bibinfo {volume} {100}},\ \bibinfo {pages}
  {245139} (\bibinfo {year} {2019})}\BibitemShut {NoStop}%
\bibitem [{sup()}]{supplemental}%
  \BibitemOpen
  \href@noop {} {}\bibinfo {note} {See Supplemental Material at [URL will be
  inserted by publisher]}\BibitemShut {NoStop}%
\bibitem [{\citenamefont {Nishizaki}\ \emph {et~al.}(1996)\citenamefont
  {Nishizaki}, \citenamefont {Maeno},\ and\ \citenamefont
  {Fujita}}]{nishizaki1996effect}%
  \BibitemOpen
  \bibfield  {author} {\bibinfo {author} {\bibfnamefont {S.}~\bibnamefont
  {Nishizaki}}, \bibinfo {author} {\bibfnamefont {Y.}~\bibnamefont {Maeno}},\
  and\ \bibinfo {author} {\bibfnamefont {T.}~\bibnamefont {Fujita}},\
  }\bibfield  {title} {\bibinfo {title} {Effect of annealing on the
  superconductivity of {Sr$_2$RuO$_4$}},\ }\href@noop {} {\bibfield  {journal}
  {\bibinfo  {journal} {J. Phys. Soc. Jpn.}\ }\textbf {\bibinfo {volume}
  {65}},\ \bibinfo {pages} {1876} (\bibinfo {year} {1996})}\BibitemShut
  {NoStop}%
\bibitem [{Note1()}]{Note1}%
  \BibitemOpen
  \bibinfo {note} {$T_{\protect \textrm {c}}$ has been shown to be relatively
  stable to changes in oxygen partial pressure annealing conditions \cite
  {nishizaki1996effect}. Therefore, the stripe features in local $T_{\protect
  \textrm {c}}$ most likely correspond to ruthenium deficiencies that
  periodically accumulate during sample growth.}\BibitemShut {Stop}%
\bibitem [{\citenamefont {Marques}\ \emph {et~al.}(2021)\citenamefont
  {Marques}, \citenamefont {Rhodes}, \citenamefont {Fittipaldi}, \citenamefont
  {Granata}, \citenamefont {Yim}, \citenamefont {Buzio}, \citenamefont {Gerbi},
  \citenamefont {Vecchione}, \citenamefont {Rost},\ and\ \citenamefont
  {Wahl}}]{marques2021magnetic}%
  \BibitemOpen
  \bibfield  {author} {\bibinfo {author} {\bibfnamefont {C.~A.}\ \bibnamefont
  {Marques}}, \bibinfo {author} {\bibfnamefont {L.~C.}\ \bibnamefont {Rhodes}},
  \bibinfo {author} {\bibfnamefont {R.}~\bibnamefont {Fittipaldi}}, \bibinfo
  {author} {\bibfnamefont {V.}~\bibnamefont {Granata}}, \bibinfo {author}
  {\bibfnamefont {C.~M.}\ \bibnamefont {Yim}}, \bibinfo {author} {\bibfnamefont
  {R.}~\bibnamefont {Buzio}}, \bibinfo {author} {\bibfnamefont
  {A.}~\bibnamefont {Gerbi}}, \bibinfo {author} {\bibfnamefont
  {A.}~\bibnamefont {Vecchione}}, \bibinfo {author} {\bibfnamefont {A.~W.}\
  \bibnamefont {Rost}},\ and\ \bibinfo {author} {\bibfnamefont
  {P.}~\bibnamefont {Wahl}},\ }\bibfield  {title} {\bibinfo {title}
  {Magnetic-field tunable intertwined checkerboard charge order and nematicity
  in the surface layer of {Sr$_2$RuO$_4$}},\ }\href@noop {} {\bibfield
  {journal} {\bibinfo  {journal} {Advanced Materials}\ }\textbf {\bibinfo
  {volume} {33}},\ \bibinfo {pages} {2100593} (\bibinfo {year}
  {2021})}\BibitemShut {NoStop}%
\bibitem [{\citenamefont {Wang}\ \emph {et~al.}(2017)\citenamefont {Wang},
  \citenamefont {Walkup}, \citenamefont {Derry}, \citenamefont {Scaffidi},
  \citenamefont {Rak}, \citenamefont {Vig}, \citenamefont {Kogar},
  \citenamefont {Zeljkovic}, \citenamefont {Husain}, \citenamefont {Santos}
  \emph {et~al.}}]{wang2017quasiparticle}%
  \BibitemOpen
  \bibfield  {author} {\bibinfo {author} {\bibfnamefont {Z.}~\bibnamefont
  {Wang}}, \bibinfo {author} {\bibfnamefont {D.}~\bibnamefont {Walkup}},
  \bibinfo {author} {\bibfnamefont {P.}~\bibnamefont {Derry}}, \bibinfo
  {author} {\bibfnamefont {T.}~\bibnamefont {Scaffidi}}, \bibinfo {author}
  {\bibfnamefont {M.}~\bibnamefont {Rak}}, \bibinfo {author} {\bibfnamefont
  {S.}~\bibnamefont {Vig}}, \bibinfo {author} {\bibfnamefont {A.}~\bibnamefont
  {Kogar}}, \bibinfo {author} {\bibfnamefont {I.}~\bibnamefont {Zeljkovic}},
  \bibinfo {author} {\bibfnamefont {A.}~\bibnamefont {Husain}}, \bibinfo
  {author} {\bibfnamefont {L.~H.}\ \bibnamefont {Santos}}, \emph {et~al.},\
  }\bibfield  {title} {\bibinfo {title} {Quasiparticle interference and strong
  electron--mode coupling in the quasi-one-dimensional bands of
  {Sr$_2$RuO$_4$}},\ }\href@noop {} {\bibfield  {journal} {\bibinfo  {journal}
  {Nat. Phys.}\ }\textbf {\bibinfo {volume} {13}},\ \bibinfo {pages} {799}
  (\bibinfo {year} {2017})}\BibitemShut {NoStop}%
\bibitem [{\citenamefont {Kambara}\ \emph {et~al.}(2006)\citenamefont
  {Kambara}, \citenamefont {Niimi}, \citenamefont {Takizawa}, \citenamefont
  {Yaguchi}, \citenamefont {Maeno},\ and\ \citenamefont
  {Fukuyama}}]{kambara2006scanning}%
  \BibitemOpen
  \bibfield  {author} {\bibinfo {author} {\bibfnamefont {H.}~\bibnamefont
  {Kambara}}, \bibinfo {author} {\bibfnamefont {Y.}~\bibnamefont {Niimi}},
  \bibinfo {author} {\bibfnamefont {K.}~\bibnamefont {Takizawa}}, \bibinfo
  {author} {\bibfnamefont {H.}~\bibnamefont {Yaguchi}}, \bibinfo {author}
  {\bibfnamefont {Y.}~\bibnamefont {Maeno}},\ and\ \bibinfo {author}
  {\bibfnamefont {H.}~\bibnamefont {Fukuyama}},\ }\bibfield  {title} {\bibinfo
  {title} {Scanning tunneling microscopy and spectroscopy of {Sr$_2$RuO$_4$}},\
  }in\ \href@noop {} {\emph {\bibinfo {booktitle} {AIP Conf. Proc.}}},\ Vol.\
  \bibinfo {volume} {850}\ (\bibinfo {year} {2006})\ pp.\ \bibinfo {pages}
  {539--540}\BibitemShut {NoStop}%
\bibitem [{\citenamefont {Barker}\ \emph {et~al.}(2003)\citenamefont {Barker},
  \citenamefont {Dutta}, \citenamefont {Lupien}, \citenamefont {McEuen},
  \citenamefont {Kikugawa}, \citenamefont {Maeno},\ and\ \citenamefont
  {Davis}}]{barker2003stm}%
  \BibitemOpen
  \bibfield  {author} {\bibinfo {author} {\bibfnamefont {B.}~\bibnamefont
  {Barker}}, \bibinfo {author} {\bibfnamefont {S.}~\bibnamefont {Dutta}},
  \bibinfo {author} {\bibfnamefont {C.}~\bibnamefont {Lupien}}, \bibinfo
  {author} {\bibfnamefont {P.}~\bibnamefont {McEuen}}, \bibinfo {author}
  {\bibfnamefont {N.}~\bibnamefont {Kikugawa}}, \bibinfo {author}
  {\bibfnamefont {Y.}~\bibnamefont {Maeno}},\ and\ \bibinfo {author}
  {\bibfnamefont {J.}~\bibnamefont {Davis}},\ }\bibfield  {title} {\bibinfo
  {title} {{STM} studies of individual {Ti} impurity atoms in
  {Sr$_2$RuO$_4$}},\ }\href@noop {} {\bibfield  {journal} {\bibinfo  {journal}
  {Physica B: Condensed Matter}\ }\textbf {\bibinfo {volume} {329}},\ \bibinfo
  {pages} {1334} (\bibinfo {year} {2003})}\BibitemShut {NoStop}%
\bibitem [{\citenamefont {Kreisel}\ \emph {et~al.}(2021)\citenamefont
  {Kreisel}, \citenamefont {Marques}, \citenamefont {Rhodes}, \citenamefont
  {Kong}, \citenamefont {Berlijn}, \citenamefont {Fittipaldi}, \citenamefont
  {Granata}, \citenamefont {Vecchione}, \citenamefont {Wahl},\ and\
  \citenamefont {Hirschfeld}}]{kreisel2021quasi}%
  \BibitemOpen
  \bibfield  {author} {\bibinfo {author} {\bibfnamefont {A.}~\bibnamefont
  {Kreisel}}, \bibinfo {author} {\bibfnamefont {C.}~\bibnamefont {Marques}},
  \bibinfo {author} {\bibfnamefont {L.}~\bibnamefont {Rhodes}}, \bibinfo
  {author} {\bibfnamefont {X.}~\bibnamefont {Kong}}, \bibinfo {author}
  {\bibfnamefont {T.}~\bibnamefont {Berlijn}}, \bibinfo {author} {\bibfnamefont
  {R.}~\bibnamefont {Fittipaldi}}, \bibinfo {author} {\bibfnamefont
  {V.}~\bibnamefont {Granata}}, \bibinfo {author} {\bibfnamefont
  {A.}~\bibnamefont {Vecchione}}, \bibinfo {author} {\bibfnamefont
  {P.}~\bibnamefont {Wahl}},\ and\ \bibinfo {author} {\bibfnamefont
  {P.}~\bibnamefont {Hirschfeld}},\ }\bibfield  {title} {\bibinfo {title}
  {Quasi-particle interference of the van hove singularity in
  {Sr$_2$RuO$_4$}},\ }\href@noop {} {\bibfield  {journal} {\bibinfo  {journal}
  {npj Quantum Mater.}\ }\textbf {\bibinfo {volume} {6}},\ \bibinfo {pages}
  {100} (\bibinfo {year} {2021})}\BibitemShut {NoStop}%
\bibitem [{\citenamefont {Matzdorf}\ \emph {et~al.}(2000)\citenamefont
  {Matzdorf}, \citenamefont {Fang}, \citenamefont {Ismail}, \citenamefont
  {Zhang}, \citenamefont {Kimura}, \citenamefont {Tokura}, \citenamefont
  {Terakura},\ and\ \citenamefont {Plummer}}]{matzdorf2000ferromagnetism}%
  \BibitemOpen
  \bibfield  {author} {\bibinfo {author} {\bibfnamefont {R.}~\bibnamefont
  {Matzdorf}}, \bibinfo {author} {\bibfnamefont {Z.}~\bibnamefont {Fang}},
  \bibinfo {author} {\bibnamefont {Ismail}}, \bibinfo {author} {\bibfnamefont
  {J.}~\bibnamefont {Zhang}}, \bibinfo {author} {\bibfnamefont
  {T.}~\bibnamefont {Kimura}}, \bibinfo {author} {\bibfnamefont
  {Y.}~\bibnamefont {Tokura}}, \bibinfo {author} {\bibfnamefont
  {K.}~\bibnamefont {Terakura}},\ and\ \bibinfo {author} {\bibfnamefont
  {E.}~\bibnamefont {Plummer}},\ }\bibfield  {title} {\bibinfo {title}
  {Ferromagnetism stabilized by lattice distortion at the surface of the p-wave
  superconductor {Sr$_2$RuO$_4$}},\ }\href@noop {} {\bibfield  {journal}
  {\bibinfo  {journal} {Science}\ }\textbf {\bibinfo {volume} {289}},\ \bibinfo
  {pages} {746} (\bibinfo {year} {2000})}\BibitemShut {NoStop}%
\bibitem [{\citenamefont {Veenstra}\ \emph {et~al.}(2013)\citenamefont
  {Veenstra}, \citenamefont {Zhu}, \citenamefont {Ludbrook}, \citenamefont
  {Capsoni}, \citenamefont {Levy}, \citenamefont {Nicolaou}, \citenamefont
  {Rosen}, \citenamefont {Comin}, \citenamefont {Kittaka}, \citenamefont
  {Maeno}, \citenamefont {Elfimov},\ and\ \citenamefont
  {Damascelli}}]{veenstra2013arpes}%
  \BibitemOpen
  \bibfield  {author} {\bibinfo {author} {\bibfnamefont {C.~N.}\ \bibnamefont
  {Veenstra}}, \bibinfo {author} {\bibfnamefont {Z.-H.}\ \bibnamefont {Zhu}},
  \bibinfo {author} {\bibfnamefont {B.}~\bibnamefont {Ludbrook}}, \bibinfo
  {author} {\bibfnamefont {M.}~\bibnamefont {Capsoni}}, \bibinfo {author}
  {\bibfnamefont {G.}~\bibnamefont {Levy}}, \bibinfo {author} {\bibfnamefont
  {A.}~\bibnamefont {Nicolaou}}, \bibinfo {author} {\bibfnamefont {J.~A.}\
  \bibnamefont {Rosen}}, \bibinfo {author} {\bibfnamefont {R.}~\bibnamefont
  {Comin}}, \bibinfo {author} {\bibfnamefont {S.}~\bibnamefont {Kittaka}},
  \bibinfo {author} {\bibfnamefont {Y.}~\bibnamefont {Maeno}}, \bibinfo
  {author} {\bibfnamefont {I.~S.}\ \bibnamefont {Elfimov}},\ and\ \bibinfo
  {author} {\bibfnamefont {A.}~\bibnamefont {Damascelli}},\ }\bibfield  {title}
  {\bibinfo {title} {Determining the surface-to-bulk progression in the
  normal-state electronic structure of {Sr$_2$RuO$_4$} by angle-resolved
  photoemission and density functional theory},\ }\href@noop {} {\bibfield
  {journal} {\bibinfo  {journal} {Phys. Rev. Lett.}\ }\textbf {\bibinfo
  {volume} {110}},\ \bibinfo {pages} {097004} (\bibinfo {year}
  {2013})}\BibitemShut {NoStop}%
\bibitem [{\citenamefont {Abarca~Morales}\ \emph {et~al.}(2023)\citenamefont
  {Abarca~Morales}, \citenamefont {Siemann}, \citenamefont {Zivanovic},
  \citenamefont {Murgatroyd}, \citenamefont {Markovi\ifmmode~\acute{c}\else
  \'{c}\fi{}}, \citenamefont {Edwards}, \citenamefont {Hooley}, \citenamefont
  {Sokolov}, \citenamefont {Kikugawa}, \citenamefont {Cacho}, \citenamefont
  {Watson}, \citenamefont {Kim}, \citenamefont {Hicks}, \citenamefont
  {Mackenzie},\ and\ \citenamefont {King}}]{morales2023hierarchy}%
  \BibitemOpen
  \bibfield  {author} {\bibinfo {author} {\bibfnamefont {E.}~\bibnamefont
  {Abarca~Morales}}, \bibinfo {author} {\bibfnamefont {G.-R.}\ \bibnamefont
  {Siemann}}, \bibinfo {author} {\bibfnamefont {A.}~\bibnamefont {Zivanovic}},
  \bibinfo {author} {\bibfnamefont {P.~A.~E.}\ \bibnamefont {Murgatroyd}},
  \bibinfo {author} {\bibfnamefont {I.}~\bibnamefont
  {Markovi\ifmmode~\acute{c}\else \'{c}\fi{}}}, \bibinfo {author}
  {\bibfnamefont {B.}~\bibnamefont {Edwards}}, \bibinfo {author} {\bibfnamefont
  {C.~A.}\ \bibnamefont {Hooley}}, \bibinfo {author} {\bibfnamefont {D.~A.}\
  \bibnamefont {Sokolov}}, \bibinfo {author} {\bibfnamefont {N.}~\bibnamefont
  {Kikugawa}}, \bibinfo {author} {\bibfnamefont {C.}~\bibnamefont {Cacho}},
  \bibinfo {author} {\bibfnamefont {M.~D.}\ \bibnamefont {Watson}}, \bibinfo
  {author} {\bibfnamefont {T.~K.}\ \bibnamefont {Kim}}, \bibinfo {author}
  {\bibfnamefont {C.~W.}\ \bibnamefont {Hicks}}, \bibinfo {author}
  {\bibfnamefont {A.~P.}\ \bibnamefont {Mackenzie}},\ and\ \bibinfo {author}
  {\bibfnamefont {P.~D.~C.}\ \bibnamefont {King}},\ }\bibfield  {title}
  {\bibinfo {title} {Hierarchy of lifshitz transitions in the surface
  electronic structure of ${\mathrm{sr}}_{2}{\mathrm{ruo}}_{4}$ under uniaxial
  compression},\ }\href@noop {} {\bibfield  {journal} {\bibinfo  {journal}
  {Phys. Rev. Lett.}\ }\textbf {\bibinfo {volume} {130}},\ \bibinfo {pages}
  {096401} (\bibinfo {year} {2023})}\BibitemShut {NoStop}%
\bibitem [{\citenamefont {Firmo}\ \emph {et~al.}(2013)\citenamefont {Firmo},
  \citenamefont {Lederer}, \citenamefont {Lupien}, \citenamefont {Mackenzie},
  \citenamefont {Davis},\ and\ \citenamefont {Kivelson}}]{firmo2013evidence}%
  \BibitemOpen
  \bibfield  {author} {\bibinfo {author} {\bibfnamefont {I.}~\bibnamefont
  {Firmo}}, \bibinfo {author} {\bibfnamefont {S.}~\bibnamefont {Lederer}},
  \bibinfo {author} {\bibfnamefont {C.}~\bibnamefont {Lupien}}, \bibinfo
  {author} {\bibfnamefont {A.~P.}\ \bibnamefont {Mackenzie}}, \bibinfo {author}
  {\bibfnamefont {J.}~\bibnamefont {Davis}},\ and\ \bibinfo {author}
  {\bibfnamefont {S.}~\bibnamefont {Kivelson}},\ }\bibfield  {title} {\bibinfo
  {title} {Evidence from tunneling spectroscopy for a quasi-one-dimensional
  origin of superconductivity in {Sr$_2$RuO$_4$}},\ }\href@noop {} {\bibfield
  {journal} {\bibinfo  {journal} {Phys. Rev. B}\ }\textbf {\bibinfo {volume}
  {88}},\ \bibinfo {pages} {134521} (\bibinfo {year} {2013})}\BibitemShut
  {NoStop}%
\bibitem [{\citenamefont {Sharma}\ \emph {et~al.}(2020)\citenamefont {Sharma},
  \citenamefont {Edkins}, \citenamefont {Wang}, \citenamefont {Kostin},
  \citenamefont {Sow}, \citenamefont {Maeno}, \citenamefont {Mackenzie},
  \citenamefont {Davis},\ and\ \citenamefont {Madhavan}}]{sharma2020momentum}%
  \BibitemOpen
  \bibfield  {author} {\bibinfo {author} {\bibfnamefont {R.}~\bibnamefont
  {Sharma}}, \bibinfo {author} {\bibfnamefont {S.~D.}\ \bibnamefont {Edkins}},
  \bibinfo {author} {\bibfnamefont {Z.}~\bibnamefont {Wang}}, \bibinfo {author}
  {\bibfnamefont {A.}~\bibnamefont {Kostin}}, \bibinfo {author} {\bibfnamefont
  {C.}~\bibnamefont {Sow}}, \bibinfo {author} {\bibfnamefont {Y.}~\bibnamefont
  {Maeno}}, \bibinfo {author} {\bibfnamefont {A.~P.}\ \bibnamefont
  {Mackenzie}}, \bibinfo {author} {\bibfnamefont {J.~S.}\ \bibnamefont
  {Davis}},\ and\ \bibinfo {author} {\bibfnamefont {V.}~\bibnamefont
  {Madhavan}},\ }\bibfield  {title} {\bibinfo {title} {Momentum-resolved
  superconducting energy gaps of {Sr$_2$RuO$_4$} from quasiparticle
  interference imaging},\ }\href@noop {} {\bibfield  {journal} {\bibinfo
  {journal} {Proceedings of the National Academy of Sciences}\ }\textbf
  {\bibinfo {volume} {117}},\ \bibinfo {pages} {5222} (\bibinfo {year}
  {2020})}\BibitemShut {NoStop}%
\bibitem [{\citenamefont {Shen}\ \emph {et~al.}(2007)\citenamefont {Shen},
  \citenamefont {Kikugawa}, \citenamefont {Bergemann}, \citenamefont {Balicas},
  \citenamefont {Baumberger}, \citenamefont {Meevasana}, \citenamefont {Ingle},
  \citenamefont {Maeno}, \citenamefont {Shen},\ and\ \citenamefont
  {Mackenzie}}]{shen2007evolution}%
  \BibitemOpen
  \bibfield  {author} {\bibinfo {author} {\bibfnamefont {K.}~\bibnamefont
  {Shen}}, \bibinfo {author} {\bibfnamefont {N.}~\bibnamefont {Kikugawa}},
  \bibinfo {author} {\bibfnamefont {C.}~\bibnamefont {Bergemann}}, \bibinfo
  {author} {\bibfnamefont {L.}~\bibnamefont {Balicas}}, \bibinfo {author}
  {\bibfnamefont {F.}~\bibnamefont {Baumberger}}, \bibinfo {author}
  {\bibfnamefont {W.}~\bibnamefont {Meevasana}}, \bibinfo {author}
  {\bibfnamefont {N.}~\bibnamefont {Ingle}}, \bibinfo {author} {\bibfnamefont
  {Y.}~\bibnamefont {Maeno}}, \bibinfo {author} {\bibfnamefont {Z.-X.}\
  \bibnamefont {Shen}},\ and\ \bibinfo {author} {\bibfnamefont {A.~P.}\
  \bibnamefont {Mackenzie}},\ }\bibfield  {title} {\bibinfo {title} {Evolution
  of the {Fermi} surface and quasiparticle renormalization through a van {Hove}
  singularity in {Sr$_{2-y}$La$_y$ RuO$_4$}},\ }\href@noop {} {\bibfield
  {journal} {\bibinfo  {journal} {Phys. Rev. Lett.}\ }\textbf {\bibinfo
  {volume} {99}},\ \bibinfo {pages} {187001} (\bibinfo {year}
  {2007})}\BibitemShut {NoStop}%
\bibitem [{\citenamefont {Li}\ \emph {et~al.}(2022)\citenamefont {Li},
  \citenamefont {Garst}, \citenamefont {Schmalian}, \citenamefont {Ghosh},
  \citenamefont {Kikugawa}, \citenamefont {Sokolov}, \citenamefont {Hicks},
  \citenamefont {Jerzembeck}, \citenamefont {Ikeda}, \citenamefont {Hu},
  \citenamefont {Ramshaw}, \citenamefont {Rost}, \citenamefont {Nicklas},\ and\
  \citenamefont {Mackenzie}}]{li2022elastocaloric}%
  \BibitemOpen
  \bibfield  {author} {\bibinfo {author} {\bibfnamefont {Y.-S.}\ \bibnamefont
  {Li}}, \bibinfo {author} {\bibfnamefont {M.}~\bibnamefont {Garst}}, \bibinfo
  {author} {\bibfnamefont {J.}~\bibnamefont {Schmalian}}, \bibinfo {author}
  {\bibfnamefont {S.}~\bibnamefont {Ghosh}}, \bibinfo {author} {\bibfnamefont
  {N.}~\bibnamefont {Kikugawa}}, \bibinfo {author} {\bibfnamefont {D.~A.}\
  \bibnamefont {Sokolov}}, \bibinfo {author} {\bibfnamefont {C.~W.}\
  \bibnamefont {Hicks}}, \bibinfo {author} {\bibfnamefont {F.}~\bibnamefont
  {Jerzembeck}}, \bibinfo {author} {\bibfnamefont {M.~S.}\ \bibnamefont
  {Ikeda}}, \bibinfo {author} {\bibfnamefont {Z.}~\bibnamefont {Hu}}, \bibinfo
  {author} {\bibfnamefont {B.~J.}\ \bibnamefont {Ramshaw}}, \bibinfo {author}
  {\bibfnamefont {A.~W.}\ \bibnamefont {Rost}}, \bibinfo {author}
  {\bibfnamefont {M.}~\bibnamefont {Nicklas}},\ and\ \bibinfo {author}
  {\bibfnamefont {A.~P.}\ \bibnamefont {Mackenzie}},\ }\bibfield  {title}
  {\bibinfo {title} {Elastocaloric determination of the phase diagram of
  {Sr$_2$RuO$_4$}},\ }\href@noop {} {\bibfield  {journal} {\bibinfo  {journal}
  {Nature}\ }\textbf {\bibinfo {volume} {607}},\ \bibinfo {pages} {276}
  (\bibinfo {year} {2022})}\BibitemShut {NoStop}%
\bibitem [{\citenamefont {Mackenzie}\ \emph {et~al.}(1998)\citenamefont
  {Mackenzie}, \citenamefont {Haselwimmer}, \citenamefont {Tyler},
  \citenamefont {Lonzarich}, \citenamefont {Mori}, \citenamefont {Nishizaki},\
  and\ \citenamefont {Maeno}}]{mackenzie1998extremely}%
  \BibitemOpen
  \bibfield  {author} {\bibinfo {author} {\bibfnamefont {A.~P.}\ \bibnamefont
  {Mackenzie}}, \bibinfo {author} {\bibfnamefont {R.}~\bibnamefont
  {Haselwimmer}}, \bibinfo {author} {\bibfnamefont {A.}~\bibnamefont {Tyler}},
  \bibinfo {author} {\bibfnamefont {G.}~\bibnamefont {Lonzarich}}, \bibinfo
  {author} {\bibfnamefont {Y.}~\bibnamefont {Mori}}, \bibinfo {author}
  {\bibfnamefont {S.}~\bibnamefont {Nishizaki}},\ and\ \bibinfo {author}
  {\bibfnamefont {Y.}~\bibnamefont {Maeno}},\ }\bibfield  {title} {\bibinfo
  {title} {Extremely strong dependence of superconductivity on disorder in
  {Sr$_2$RuO$_4$}},\ }\href@noop {} {\bibfield  {journal} {\bibinfo  {journal}
  {Phys. Rev. Lett.}\ }\textbf {\bibinfo {volume} {80}},\ \bibinfo {pages}
  {161} (\bibinfo {year} {1998})}\BibitemShut {NoStop}%
\bibitem [{\citenamefont {Mao}\ \emph {et~al.}(1999)\citenamefont {Mao},
  \citenamefont {Mori},\ and\ \citenamefont {Maeno}}]{mao1999suppression}%
  \BibitemOpen
  \bibfield  {author} {\bibinfo {author} {\bibfnamefont {Z.}~\bibnamefont
  {Mao}}, \bibinfo {author} {\bibfnamefont {Y.}~\bibnamefont {Mori}},\ and\
  \bibinfo {author} {\bibfnamefont {Y.}~\bibnamefont {Maeno}},\ }\bibfield
  {title} {\bibinfo {title} {Suppression of superconductivity in
  {Sr$_2$RuO$_4$} caused by defects},\ }\href@noop {} {\bibfield  {journal}
  {\bibinfo  {journal} {Phys. Rev. B}\ }\textbf {\bibinfo {volume} {60}},\
  \bibinfo {pages} {610} (\bibinfo {year} {1999})}\BibitemShut {NoStop}%
\bibitem [{\citenamefont {Kikugawa}\ and\ \citenamefont
  {Maeno}(2002)}]{kikugawa2002non}%
  \BibitemOpen
  \bibfield  {author} {\bibinfo {author} {\bibfnamefont {N.}~\bibnamefont
  {Kikugawa}}\ and\ \bibinfo {author} {\bibfnamefont {Y.}~\bibnamefont
  {Maeno}},\ }\bibfield  {title} {\bibinfo {title} {{Non-Fermi}-liquid behavior
  in {Sr$_2$RuO$_4$} with nonmagnetic impurities},\ }\href@noop {} {\bibfield
  {journal} {\bibinfo  {journal} {Phys. Rev. Lett.}\ }\textbf {\bibinfo
  {volume} {89}},\ \bibinfo {pages} {117001} (\bibinfo {year}
  {2002})}\BibitemShut {NoStop}%
\bibitem [{\citenamefont {Kikugawa}\ \emph {et~al.}(2004)\citenamefont
  {Kikugawa}, \citenamefont {Mackenzie}, \citenamefont {Bergemann},
  \citenamefont {Borzi}, \citenamefont {Grigera},\ and\ \citenamefont
  {Maeno}}]{kikugawa2004rigid}%
  \BibitemOpen
  \bibfield  {author} {\bibinfo {author} {\bibfnamefont {N.}~\bibnamefont
  {Kikugawa}}, \bibinfo {author} {\bibfnamefont {A.~P.}\ \bibnamefont
  {Mackenzie}}, \bibinfo {author} {\bibfnamefont {C.}~\bibnamefont
  {Bergemann}}, \bibinfo {author} {\bibfnamefont {R.}~\bibnamefont {Borzi}},
  \bibinfo {author} {\bibfnamefont {S.}~\bibnamefont {Grigera}},\ and\ \bibinfo
  {author} {\bibfnamefont {Y.}~\bibnamefont {Maeno}},\ }\bibfield  {title}
  {\bibinfo {title} {Rigid-band shift of the {Fermi} level in the strongly
  correlated metal: {Sr$_{2-y}$La$_y$RuO$_4$}},\ }\href@noop {} {\bibfield
  {journal} {\bibinfo  {journal} {Phys. Rev. B}\ }\textbf {\bibinfo {volume}
  {70}},\ \bibinfo {pages} {060508} (\bibinfo {year} {2004})}\BibitemShut
  {NoStop}%
\bibitem [{\citenamefont {Abrikosov}\ and\ \citenamefont
  {Gor'kov}(1960)}]{abrikosov1960contribution}%
  \BibitemOpen
  \bibfield  {author} {\bibinfo {author} {\bibfnamefont {A.~A.}\ \bibnamefont
  {Abrikosov}}\ and\ \bibinfo {author} {\bibfnamefont {L.~P.}\ \bibnamefont
  {Gor'kov}},\ }\bibfield  {title} {\bibinfo {title} {Contribution to the
  theory of superconducting alloys with paramagnetic impurities},\ }\href@noop
  {} {\bibfield  {journal} {\bibinfo  {journal} {Zhur. Eksptl'. i Teoret.
  Fiz.}\ }\textbf {\bibinfo {volume} {39}} (\bibinfo {year}
  {1960})}\BibitemShut {NoStop}%
\bibitem [{\citenamefont {Hirschfeld}\ \emph {et~al.}(1988)\citenamefont
  {Hirschfeld}, \citenamefont {W{\"o}lfle},\ and\ \citenamefont
  {Einzel}}]{hirschfeld1988consequences}%
  \BibitemOpen
  \bibfield  {author} {\bibinfo {author} {\bibfnamefont {P.}~\bibnamefont
  {Hirschfeld}}, \bibinfo {author} {\bibfnamefont {P.}~\bibnamefont
  {W{\"o}lfle}},\ and\ \bibinfo {author} {\bibfnamefont {D.}~\bibnamefont
  {Einzel}},\ }\bibfield  {title} {\bibinfo {title} {Consequences of resonant
  impurity scattering in anisotropic superconductors: Thermal and spin
  relaxation properties},\ }\href@noop {} {\bibfield  {journal} {\bibinfo
  {journal} {Phys. Rev. B}\ }\textbf {\bibinfo {volume} {37}},\ \bibinfo
  {pages} {83} (\bibinfo {year} {1988})}\BibitemShut {NoStop}%
\bibitem [{\citenamefont {Landaeta}\ \emph {et~al.}(2023)\citenamefont
  {Landaeta}, \citenamefont {Semeniuk}, \citenamefont {Aretz}, \citenamefont
  {Shirer}, \citenamefont {Sokolov}, \citenamefont {Kikugawa}, \citenamefont
  {Maeno}, \citenamefont {Bonalde}, \citenamefont {Schmalian}, \citenamefont
  {Mackenzie},\ and\ \citenamefont {Hassinger}}]{landaeta2023nonlocal}%
  \BibitemOpen
  \bibfield  {author} {\bibinfo {author} {\bibfnamefont {J.~F.}\ \bibnamefont
  {Landaeta}}, \bibinfo {author} {\bibfnamefont {K.}~\bibnamefont {Semeniuk}},
  \bibinfo {author} {\bibfnamefont {J.}~\bibnamefont {Aretz}}, \bibinfo
  {author} {\bibfnamefont {K.}~\bibnamefont {Shirer}}, \bibinfo {author}
  {\bibfnamefont {D.~A.}\ \bibnamefont {Sokolov}}, \bibinfo {author}
  {\bibfnamefont {N.}~\bibnamefont {Kikugawa}}, \bibinfo {author}
  {\bibfnamefont {Y.}~\bibnamefont {Maeno}}, \bibinfo {author} {\bibfnamefont
  {I.}~\bibnamefont {Bonalde}}, \bibinfo {author} {\bibfnamefont
  {J.}~\bibnamefont {Schmalian}}, \bibinfo {author} {\bibfnamefont {A.~P.}\
  \bibnamefont {Mackenzie}},\ and\ \bibinfo {author} {\bibfnamefont
  {E.}~\bibnamefont {Hassinger}},\ }\bibfield  {title} {\bibinfo {title}
  {Evidence for vertical line nodes in {Sr$_2$RuO$_4$} from nonlocal
  electrodynamics},\ }\href@noop {} {\bibfield  {journal} {\bibinfo  {journal}
  {arXiv preprint arXiv:2312.05129}\ } (\bibinfo {year} {2023})}\BibitemShut
  {NoStop}%
\end{thebibliography}%


\begin{thebibliography}{2}%
\makeatletter
\providecommand \@ifxundefined [1]{%
 \@ifx{#1\undefined}
}%
\providecommand \@ifnum [1]{%
 \ifnum #1\expandafter \@firstoftwo
 \else \expandafter \@secondoftwo
 \fi
}%
\providecommand \@ifx [1]{%
 \ifx #1\expandafter \@firstoftwo
 \else \expandafter \@secondoftwo
 \fi
}%
\providecommand \natexlab [1]{#1}%
\providecommand \enquote  [1]{``#1''}%
\providecommand \bibnamefont  [1]{#1}%
\providecommand \bibfnamefont [1]{#1}%
\providecommand \citenamefont [1]{#1}%
\providecommand \href@noop [0]{\@secondoftwo}%
\providecommand \href [0]{\begingroup \@sanitize@url \@href}%
\providecommand \@href[1]{\@@startlink{#1}\@@href}%
\providecommand \@@href[1]{\endgroup#1\@@endlink}%
\providecommand \@sanitize@url [0]{\catcode `\\12\catcode `\$12\catcode
  `\&12\catcode `\#12\catcode `\^12\catcode `\_12\catcode `\%12\relax}%
\providecommand \@@startlink[1]{}%
\providecommand \@@endlink[0]{}%
\providecommand \url  [0]{\begingroup\@sanitize@url \@url }%
\providecommand \@url [1]{\endgroup\@href {#1}{\urlprefix }}%
\providecommand \urlprefix  [0]{URL }%
\providecommand \Eprint [0]{\href }%
\providecommand \doibase [0]{https://doi.org/}%
\providecommand \selectlanguage [0]{\@gobble}%
\providecommand \bibinfo  [0]{\@secondoftwo}%
\providecommand \bibfield  [0]{\@secondoftwo}%
\providecommand \translation [1]{[#1]}%
\providecommand \BibitemOpen [0]{}%
\providecommand \bibitemStop [0]{}%
\providecommand \bibitemNoStop [0]{.\EOS\space}%
\providecommand \EOS [0]{\spacefactor3000\relax}%
\providecommand \BibitemShut  [1]{\csname bibitem#1\endcsname}%
\let\auto@bib@innerbib\@empty
\bibitem [{\citenamefont {Olivares~Rodriguez}(2022)}]{olivares2022stress}%
  \BibitemOpen
  \bibfield  {author} {\bibinfo {author} {\bibfnamefont {J.}~\bibnamefont
  {Olivares~Rodriguez}},\ }\emph {\bibinfo {title} {Stress and crystal
  imperfections: Tools for the exploration of unconventional superconductivity
  via scanning tunneling microscopy}},\ \href@noop {} {Ph.D. thesis},\ \bibinfo
   {school} {University of Illinois Urbana-Champaign} (\bibinfo {year}
  {2022})\BibitemShut {NoStop}%
\bibitem [{\citenamefont {Sunko}\ \emph {et~al.}(2019)\citenamefont {Sunko},
  \citenamefont {Abarca~Morales}, \citenamefont {Markovi{\'c}}, \citenamefont
  {Barber}, \citenamefont {Milosavljevi{\'c}}, \citenamefont {Mazzola},
  \citenamefont {Sokolov}, \citenamefont {Kikugawa}, \citenamefont {Cacho},
  \citenamefont {Dudin} \emph {et~al.}}]{sunko2019direct}%
  \BibitemOpen
  \bibfield  {author} {\bibinfo {author} {\bibfnamefont {V.}~\bibnamefont
  {Sunko}}, \bibinfo {author} {\bibfnamefont {E.}~\bibnamefont
  {Abarca~Morales}}, \bibinfo {author} {\bibfnamefont {I.}~\bibnamefont
  {Markovi{\'c}}}, \bibinfo {author} {\bibfnamefont {M.~E.}\ \bibnamefont
  {Barber}}, \bibinfo {author} {\bibfnamefont {D.}~\bibnamefont
  {Milosavljevi{\'c}}}, \bibinfo {author} {\bibfnamefont {F.}~\bibnamefont
  {Mazzola}}, \bibinfo {author} {\bibfnamefont {D.~A.}\ \bibnamefont
  {Sokolov}}, \bibinfo {author} {\bibfnamefont {N.}~\bibnamefont {Kikugawa}},
  \bibinfo {author} {\bibfnamefont {C.}~\bibnamefont {Cacho}}, \bibinfo
  {author} {\bibfnamefont {P.}~\bibnamefont {Dudin}}, \emph {et~al.},\
  }\bibfield  {title} {\bibinfo {title} {Direct observation of a uniaxial
  stress-driven {Lifshitz} transition in {Sr$_2$RuO$_4$}},\ }\href@noop {}
  {\bibfield  {journal} {\bibinfo  {journal} {npj Quantum Mater.}\ }\textbf
  {\bibinfo {volume} {4}},\ \bibinfo {pages} {46} (\bibinfo {year}
  {2019})}\BibitemShut {NoStop}%
\end{thebibliography}%

\end{document}


\title{Supplementary Materials for: Superconducting Penetration Depth Through a Van Hove Singularity: \texorpdfstring{Sr\textsubscript{2}}\texorpdfstring{RuO\textsubscript{4}} Under Uniaxial Stress}

\author{Eli Mueller}
\affiliation{Stanford Institute for Materials and Energy Sciences, SLAC National Accelerator Laboratory, 2575 Sand Hill Road, Menlo Park, California 94025, USA}
\affiliation{Department of Physics, Stanford University, Stanford, California 94305, USA}

\author{Yusuke Iguchi}

\affiliation{Stanford Institute for Materials and Energy Sciences, SLAC National Accelerator Laboratory, 2575 Sand Hill Road, Menlo Park, California 94025, USA}
\affiliation{Geballe Laboratory for Advanced Materials, Stanford University, Stanford, California 94305, USA}

\author{Fabian Jerzembeck}
\affiliation{Max Planck Institute for the Chemical Physics of Solids, N{\"o}thnitzer Stra{\ss}e 40, Dresden 01187, Germany}

\author{Jorge O. Rodriguez}
\affiliation{Department of Physics, University of Illinois, Urbana, Illinois 61801, USA}

\author{Marisa Romanelli}
\affiliation{Department of Physics, University of Illinois, Urbana, Illinois 61801, USA}

\author{Edgar Abarca-Morales}
\affiliation{Max Planck Institute for the Chemical Physics of Solids, N{\"o}thnitzer Stra{\ss}e 40, Dresden 01187, Germany}

\author{Anastasios Markou}
\affiliation{Max Planck Institute for the Chemical Physics of Solids, N{\"o}thnitzer Stra{\ss}e 40, Dresden 01187, Germany}
\affiliation{Department of Physics, University of Ioannina, 45110 Ioannina, Greece}

\author{Naoki Kikugawa}
\affiliation{National Institute for Materials Science, Tsukuba, Ibaraki 305-0003, Japan}

\author{Dmitry A. Sokolov}
\affiliation{Max Planck Institute for the Chemical Physics of Solids, N{\"o}thnitzer Stra{\ss}e 40, Dresden 01187, Germany}

\author{Gwansuk Oh}
\affiliation{Department of Physics, Graduate School of Science, Kyoto University, Kyoto 606-8502, Japan}
\affiliation{Department of Physics, Pohang University of Science and Technology (POSTECH),  Pohang 790-784, Republic of Korea}

\author{Clifford W. Hicks}
\affiliation{Max Planck Institute for the Chemical Physics of Solids, N{\"o}thnitzer Stra{\ss}e 40, Dresden 01187, Germany}
\affiliation{School of Physics and Astronomy, University of Birmingham, Birmingham B15 2TT, United Kingdom}

\author{Andrew P. Mackenzie}
\affiliation{Max Planck Institute for the Chemical Physics of Solids, N{\"o}thnitzer Stra{\ss}e 40, Dresden 01187, Germany}
\affiliation{Scottish Universities Physics Alliance, School of Physics and Astronomy, University of St. Andrews,St. Andrews KY16 9SS, United Kingdom}

\author{Yoshiteru Maeno}
\affiliation{Department of Physics, Graduate School of Science, Kyoto University, Kyoto 606-8502, Japan}
\affiliation{Toyota Riken - Kyoto University Research Center (TRiKUC), Kyoto University, Kyoto 606-8501, Japan}

\author{Vidya Madhavan}
\affiliation{Department of Physics, University of Illinois, Urbana, Illinois 61801, USA}

\author{Kathryn A. Moler}
\affiliation{Stanford Institute for Materials and Energy Sciences, SLAC National Accelerator Laboratory, 2575 Sand Hill Road, Menlo Park, California 94025, USA}
\affiliation{Department of Physics, Stanford University, Stanford, California 94305, USA}
\affiliation{Geballe Laboratory for Advanced Materials, Stanford University, Stanford, California 94305, USA}

 \maketitle

\section{Additional information: STM}

 Uniaxial strain was applied in the STM measurement using a thermal expansion cell described in Ref.~\cite{olivares2022stress} and adapted from that used in Ref.~\cite{sunko2019direct}. The STM strain cell was calibrated using digital image correlation (DIC) and elastoresistivity measurements described in Ref. \cite{olivares2022stress} which estimated the strain of the sample platform to be approximately -0.47\% to -0.5\%. The surface of the sample is under less strain than the sample platform itself and based on profilometry measurements of the sample thickness after cleaving, we estimate the strain at the sample surface to be approximately -0.4\%.

\begin{figure}[h]
 \centerline{\includegraphics[width=\linewidth]{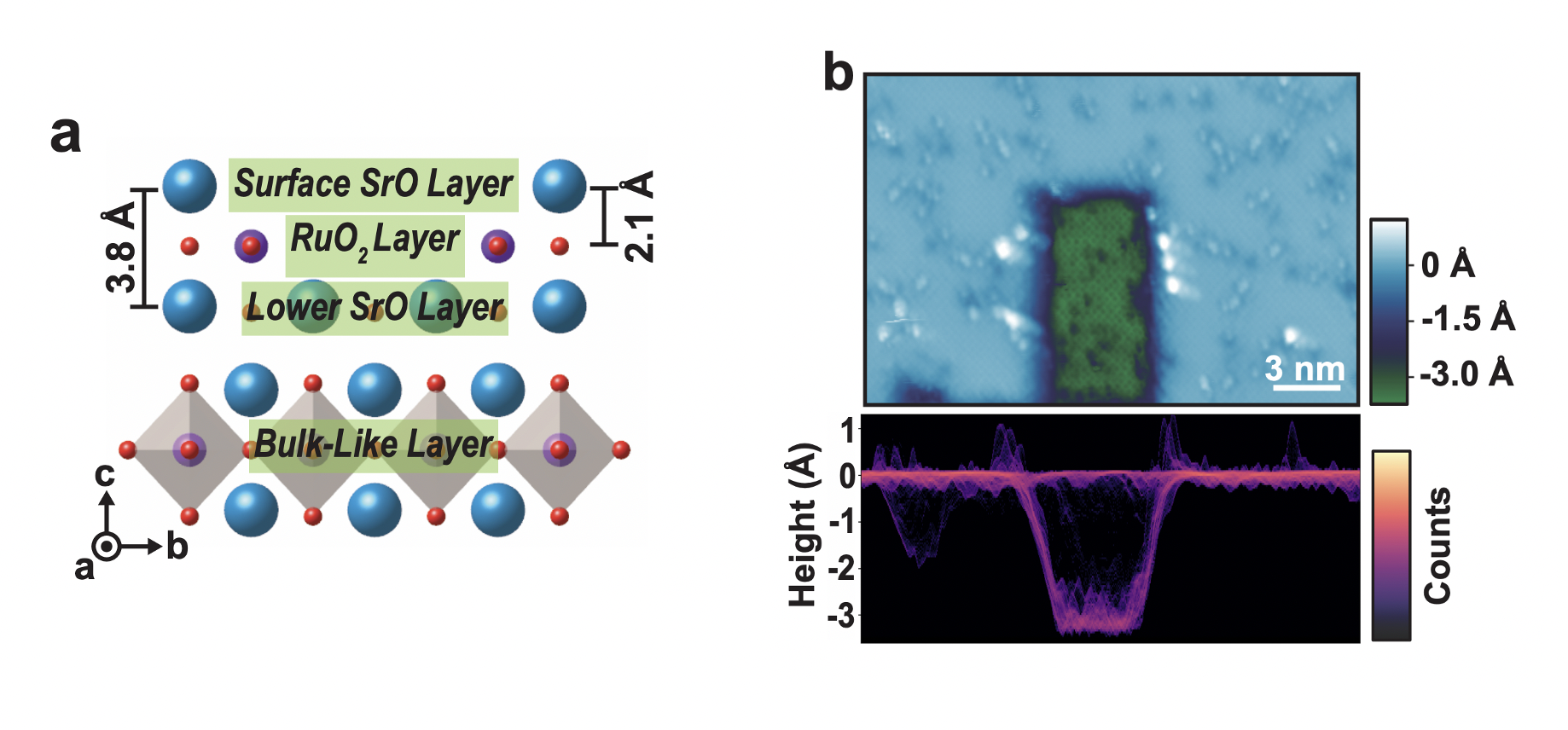}}
  \caption{(a) Schematic diagram of the large crystallographic defects on the surface of strontium ruthenate. We identify two different types, which expose a different crystal layer, SrO and RuO$_2$. (b) Topographic image (upper panel) of a defect on the surface of strontium ruthenate. The density histogram (lower panel) of the topographic image shown, projected along y and binned, allows us to better visualize the mean depth of 3.4~\r{A} }
  \label{fig:STM_supplemental_fig}
\end{figure}

We determined the topographical depths of the cleavage defects by flattening the three dimensional topographic dataset into two dimensions, via $T(x,y) \rightarrow T(x)$, along the vertical dimension. The topographic values on this restructured array are then binned by frequency for a particular value of $x$. The lower panel of Fig. \ref{fig:STM_supplemental_fig}(b) shows this projected and binned data, while the upper panel shows the original topography. From this analysis, we see that the difference in height between the upper and lower surfaces is approximately 3.4~\r{A}, in good agreement with the established value of 3.8 ~\r{A} for the separation between two SrO layers and much larger than the 2.1~\r{A} separation between SrO and RuO$_2$ layers, as illustrated in the schematic of the defect in Fig. \ref{fig:STM_supplemental_fig}(a).


\section{Additional information: Scanning SQUID}

The local $T_{\textrm{c}}$ maps shown in Fig.~2 of the main text were made by taking spatial scans of $M$ at various temperatures near the unstressed bulk $T_{\textrm{c}}$ of the sample and at the base temperature $T_{\textrm{base}}\approx$500~mK with a fixed scan height. At each temperature, the maps of $M(T)$ are converted to maps of $\lambda(T) + z_0$.  We assume constant $z_0$ for all scans so that temperature dependent changes in $\lambda(T) + z_0$ correspond to changes in $\lambda(T)$. For each pixel in the temperature series scans, we set $\Delta\lambda(T) = \lambda(T) - \lambda(T_{\textrm{base}})$ and we calculated the normalized superfluid density $\rho(T) = \lambda_0^2/(\lambda_0+\Delta\lambda(T))^2$ using $\lambda_0 = 190$~nm.  We performed a $T$-linear fit of $\rho(T)$ near the bulk $T_{\textrm{c}}$ and set the local $T_{\textrm{c}}$ to be the temperature of the $\rho = 0$ intercept. We note that the calculation of $\rho(T)$ here assumes a constant and spatially uniform penetration depth of $\lambda_0 = 190$~nm for $0<T<T_{\textrm{base}}$; however, varying $\lambda_0
$ by $\pm$20~nm does not substantially change the values of $\rho(T)$ very close to the bulk $T_{\textrm{c}}$ and therefore does not significantly change the resulting local $T_{\textrm{c}}$ map.

\begin{figure}[H]
 \centerline{\includegraphics[width=\linewidth]{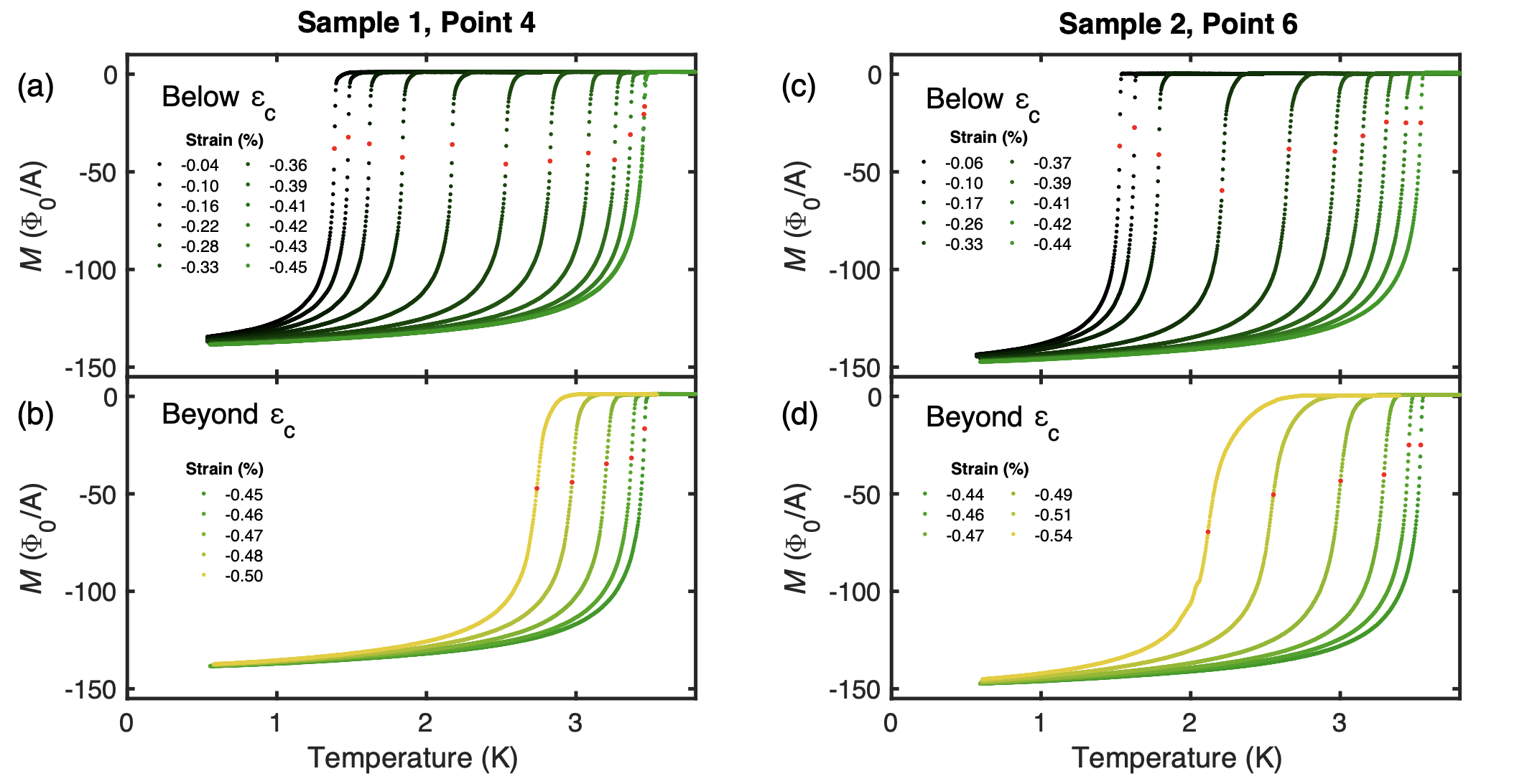}}
  \caption{(a) and (b) $M(T)$ measured on sample 1 for strains below $\varepsilon_{\textrm{c}}$ and above $\varepsilon_{\textrm{c}}$ respectively. (c) and (d) $M(T)$ measured on sample 2 for strains below $\varepsilon_{\textrm{c}}$ and above $\varepsilon_{\textrm{c}}$ respectively. Red data points indicate where the slope of $M(T)$ is maximal and the temperature at which we take to be the local $T_{\textrm{c}}$}
  \label{fig:M_vs_T}
\end{figure}

Data on the strain- and temperature-dependent changes of the penetration depth shown in Fig.~3(a) and (b) of the main text were obtained by bringing the SQUID susceptometer into mechanical contact with a selected point on the sample and recording $M$ while sweeping the temperature from $T_{\textrm{base}}\approx$500~mK to above $T_{\textrm{c}}$ for a series of applied strains. Figure \ref{fig:M_vs_T} shows the temperature dependence of the mutual inductance $M(T)$ measured under a series of compressive strains through the Lifshitz transition on sample 1 [Fig.~\ref{fig:M_vs_T}(a) and (b)] and sample 2 [Fig.~\ref{fig:M_vs_T}(c) and (d)]. We observed rounding of the superconducting transition under increasing strain due to strain inhomogeneity within the measurement volume. The local $T_{\textrm{c}}$ at each strain is defined by the temperature at which $dM/dT$ was maximal and is indicated by the red data points in \ref{fig:M_vs_T}.

\begin{figure}[H]
 \centerline{\includegraphics[width=.5\linewidth]{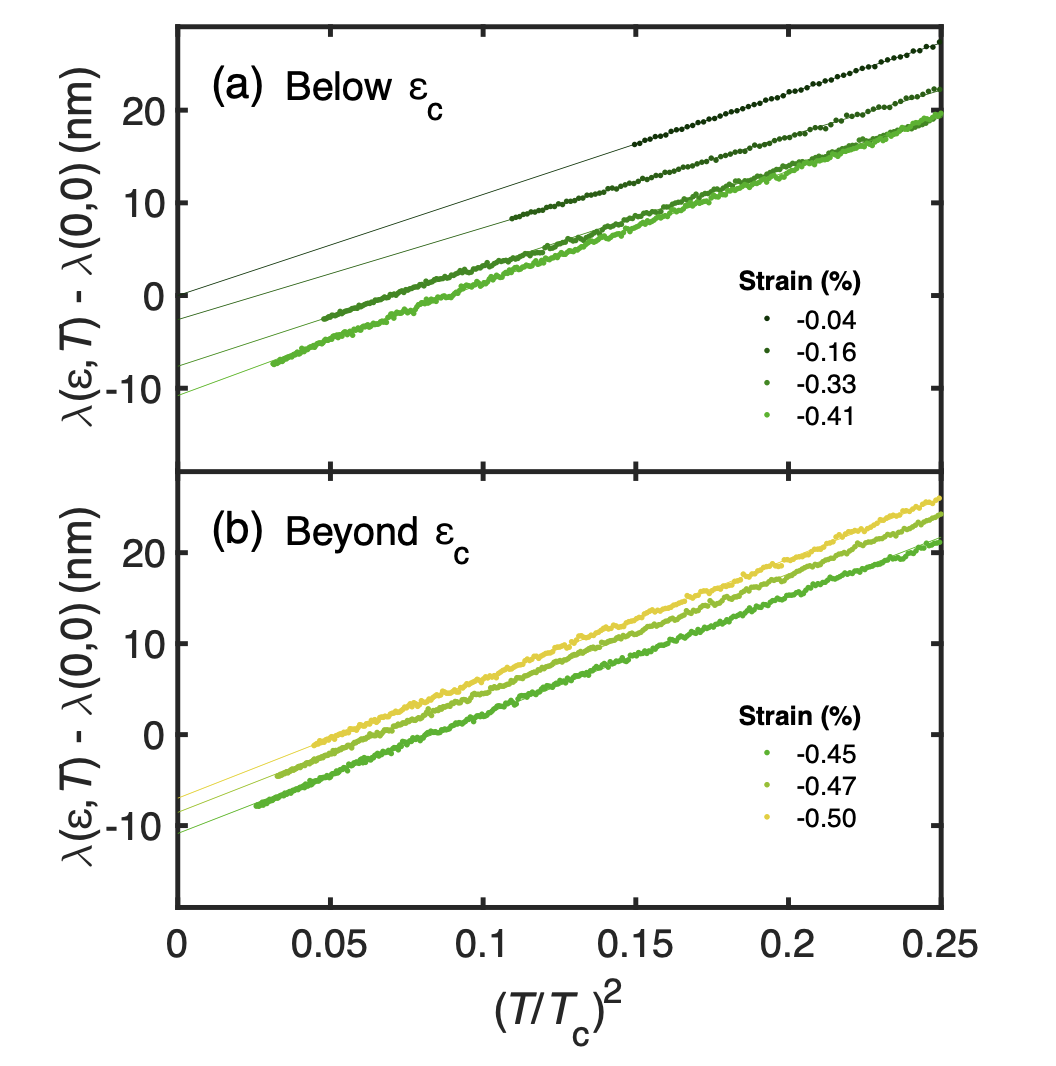}}
  \caption{Temperature and strain dependence of the penetration depth measured at point 4 on sample 1. Panels (a) and (b) correspond to strain values below and above $\varepsilon_{\textrm{c}}$ respectively.}
  \label{fig:lambda_vs_T_S1}
\end{figure}

The main text shows data on the temperature and strain dependence of the change in penetration depth $\lambda(\varepsilon,T)-\lambda(0,0)$ measured on sample 2. The data measured on sample 1 are shown in Fig. \ref{fig:lambda_vs_T_S1}. Consistent with measurements on sample 2, the penetration depth shows a $\sim T^2$ dependence for $T<0.5T_{\textrm{c}}$ thoroughout the Lifshitz transition.

\bibliography{SRO}